\documentclass[%
 reprint,
superscriptaddress,
 amsmath,amssymb,
 aps,
pra,
]{revtex4-2}
\usepackage{graphicx}
\usepackage{dcolumn}
\usepackage{bm}
\usepackage{hyperref}
\usepackage{amsfonts}
\usepackage{xcolor}
\usepackage{comment}
\usepackage{booktabs}
\usepackage{subfigure,subcaption}
\usepackage{makecell}

\begin{document}

\preprint{APS/123-QED}

\title{Experimental demonstration of the Bell-type inequalities for four qubit Dicke state using IBM Quantum Processing Units}

\author{Tomis Prajapati}
  \email{tomis.1@iitj.ac.in} 
  \affiliation{%
 Department of Physics, Indian Institute of Technology Jodhpur 342030}%
\author{Harsh Mehta}%
 \email{m23iqt011@iitj.ac.in}
 \affiliation{%
 Quantum Information and Computation (IDRP), Indian Institute of Technology Jodhpur 342030}
\author{Shreya Banerjee}
\email{shreya93ban@gmail.com}
\affiliation{Center for Quantum Science and Technology, Siksha ’O’ Anusandhan University, Bhubaneswar-751030, Odisha, India}
\author{Prasanta K. Panigrahi}
\email{pprasanta@iiserkol.ac.in}
\affiliation{Center for Quantum Science and Technology, Siksha ’O’ Anusandhan University, Bhubaneswar-751030, Odisha, India}
\affiliation{Department of Physical Sciences, Indian Institute of Science Education and Research Kolkata, Mohanpur-741246, West Bengal, India}

\author{V. Narayanan}
\email{vnara@iitj.ac.in}
 \affiliation{%
 Department of Physics, Indian Institute of Technology Jodhpur 342030}%
\affiliation{%
 Quantum Information and Computation (IDRP), Indian Institute of Technology Jodhpur 342030}
\altaffiliation{}

\begin{abstract}
Violation of the Bell-type inequalities is necessary to confirm the existence of nonlocality in nonclassical (entangled) states. We have designed a customized operator which is made of the sum of the Pauli matrices ($\sigma_x$, $\sigma_y$, and $\sigma_z$). We theoretically and experimentally investigate the violation of Bell-type inequalities using two- and four-qubit Dicke states on IBM Quantum Processing Units (QPUs). We compare two different state preparation methods for the four-qubit Dicke state—gate-based and statevector-based—and evaluate their performance on two IBM QPUs, \texttt{ibm\_kyiv} and \texttt{ibm\_sherbrook}. For the two-qubit case, we demonstrate clear violations of the CHSH inequality, with the highest observed Bell parameter reaching $2.821 \pm 0.0019$ using M3 error mitigation, which is within $0.7\sigma$ of the theoretical maximum $2\sqrt{2}$. In the four-qubit case, we employ a Bell-type inequality tailored for Dicke states and achieve a maximum violation of $2.607 \pm 0.029$ without the need for additional mitigation when using the statevector-based method. Our results reveal that advanced error mitigation techniques significantly enhance the observed violations in the gate-based method, while the statevector-based approach inherently yields more robust states with lower noise. This study highlights the critical role of state preparation and mitigation techniques in probing fundamental quantum correlations on near-term quantum hardware.
\end{abstract}


\maketitle

\section{Introduction}
Quantum entanglement is an essential phenomenon in quantum information because of its correlation between two (or many) particles \cite{nielsen_chuang_2010}. Quantum entanglement is a key component in many quantum information processing techniques,     such as quantum teleportation \cite{PhysRevLett.70.1895, PhysRevLett.69.2881}, quantum dense coding \cite{PhysRevA.59.1829,PhysRevA.61.042311}, quantum cryptography \cite{BENNETT20147,10.1119/1.16243}. Regarding quantum entanglement, the two-qubit (two-partite) Bell state has received much attention. Over the decades, extensive research has been devoted to exploring and demonstrating violations of the Bell inequality \cite{PhysRevLett.28.938, PhysRevLett.49.91, Hensen2015,PhysRevLett.115.250401, PhysRevLett.119.010402}. Still, the multipartite entangled state and its applications are also gaining investigators' attention. High-quality multipartite entangled states, such as Greenberger-Horne-Zeilinger (GHZ) states \cite{10.1119/1.16243}, W states, Cluster states, and Dicke states, are very important because they play a vital role in quantum information theory and its applications. Therefore, the efficient way of experimental generation of these states is very challenging and violations of local hidden variable theory for multipartite entangled states are also critical for certain device-independent (DI) quantum information processing tasks, including ensuring the security of quantum key distribution (QKD) \cite{PhysRevLett.67.661, PhysRevLett.87.117901}, reducing computational complexity \cite{PhysRevA.68.032309, PhysRevLett.92.127901}, and validating random number generation \cite{Pironio2010}. Detecting violations of local realism in multipartite correlations presents significant challenges, primarily due to the absence of straightforward Bell inequalities for multipartite systems. In the case of two-qubit systems, the Clauser-Horne-Shimony-Holt (CHSH) inequality \cite{PhysRevLett.23.880} is relatively simple in form, which facilitates the experimental observation of violations of local hidden variable theories. For systems involving any number of observers,  a selection between two local dichotomic observables is chosen for which researchers have developed a comprehensive set of CHSH Bell type of inequalities \cite{PhysRevA.64.032112, PhysRevA.64.010102, PhysRevLett.88.210401}. These inequalities exhibit a unified structure \cite{PhysRevA.79.022110}. Despite this, the explicit forms of Bell inequalities are frequently too intricate to effectively confirm violations of local realism. The multipartite Mermin-Ardehali-Belinskii-Klyshko (MABK) inequalities do provide explicit expressions; however, for a $N$-partite system, these inequalities require consideration of $2N$ terms \cite{PhysRevLett.65.1838, PhysRevA.46.5375, AVBelinskiĭ_1993}. The complexity of MABK inequalities makes them less practical for testing the nonlocality of multipartite entangled states, particularly as the number of particles increases. However, multipartite Bell inequalities that focus solely on two-body correlations can still detect Bell nonlocality in many-body systems. This approach provides a viable pathway for experimentally verifying multipartite Bell nonlocality \cite{tura2014detecting, tura2015nonlocality}.

Reducing the number of terms in a Bell inequality simplifies the demonstration of nonlocality for multipartite entangled states. Through an analysis of the geometrical properties of correlation polytopes, researchers have uncovered structural characteristics of CHSH-type Bell inequalities \cite{PhysRevA.88.022126}. Additionally, multipartite CHSH-type Bell inequalities with few terms have been proposed, facilitating the experimental approaches to detect violations of local realism in multipartite entangled states.
Dicke states, which naturally occur in symmetric systems \cite{PhysRev.93.99}, exhibit strong resilience against photon loss and projective measurements \cite{PhysRevLett.98.063604} and have a high degree of multipartite entanglement. Researchers have observed that Dicke states are ideal for quantum metrology \cite{PhysRevLett.107.080504, PhysRevA.85.022321} and are useful for quantifying the entanglement depth in many-body systems \cite{PhysRevLett.107.180502}. These favorable properties have spurred considerable efforts to produce high-quality Dicke states. For instance, in optical systems, a four-photon symmetric Dicke state with a fidelity of $0.844 \pm 0.008$ was experimentally realized by N. Kiesel et al. in 2007 \cite{PhysRevLett.98.063604}, and $0.904 \pm 0.004$ by Yuan Yuan Zhao et al. in 2015 \cite{Zhao:15} while the six-photon version was concurrently achieved by W. Wieczorek et al. \cite{PhysRevLett.103.020504} and R. Prevedel et al. \cite{PhysRevLett.103.020503}. However, conventional methods for detecting the entanglement of Dicke states do not always indicate violations of local realism \cite{PhysRevLett.98.063604, PhysRevLett.107.080504, PhysRevLett.103.020504, PhysRevLett.103.020503, PhysRevLett.101.010503}.

In this work, we discuss the generation of the four-qubit Dicke state and the violation of the Bell-type inequality \cite{PhysRevLett.100.200407} using our custom operators made of the sum of Pauli matrices.
To violate the Bell-type inequality for bipartite and multipartite entangled states, a specific basis for the corresponding state and inequality is required. Our customized operator is basis-independent and can be directly employed to determine the violation of the Bell inequality for both two-qubit and multi-qubit systems.

To implement the custom operators on IBM QPU, we decompose them into Pauli matrices ($\sigma_x$, $\sigma_y$, and $\sigma_z$)  with appropriate coefficients. This decomposed representation can then be implemented directly on the IBM Quantum Processing Units (QPUs). To test and simulate quantum algorithms, IBM Quantum has made its QPUs available to the public since 2016. IBM Quantum provides cloud-based access to QPUs that enable users to design and test quantum circuits on both simulators and real QPUs. IBM's Composer is a cloud-based platform for quantum computing, accessible through their website. It allows users to build quantum circuits which then can be executed either on a simulator or a quantum processing unit (QPU) via a cloud interface:( https://quantum.ibm.com/). Numerous quantum chip experiments have now been conducted using IBM QPUs. Quantum simulation \cite{unknown, Halder2018DigitalQS}, the creation of quantum algorithms \cite{GarciaMartin2018}, the testing of theoretical problems involving quantum information \cite{PhysRevA.94.012314}, quantum cryptography \cite{majumder2017experimentalrealizationsecuremultiparty}, quantum error correction \cite{Roffe_2018}, and quantum applications \cite{Mahanti2019} are just a few examples of actual experiments. Initially, we evaluated the Bell violation for the two-qubit Bell state using our customized operator on two different IBM QPUs named \texttt{kyiv}, and \texttt{sherbrook}, as well as on a noisy simulator using noise models of \texttt{ibm\_kyiv} and \texttt{ibm\_sherbrooke}, which replicate the features and noise characteristics of the actual IBM quantum devices\cite{PhysRevA.89.042123} to complement the real hardware runs we executed. Subsequently, a four-qubit Dicke state was prepared using two different methods, and its local realism was investigated by incorporating the methodology mentioned earlier.

The remainder of this article is organized as follows: in Sec. II, we describe the two distinct methods for generating the Dicke state. The development of our customized operator is outlined in Sec. III. Experimental implementation is discussed in Sec. IV. Our findings are presented in Sec. V, and we conclude the article in Sec. VI.
\section{Generation of four-qubit Dicke state}
A four-qubit Dicke state is a specific type of quantum state involving four qubits that exhibit a particular entanglement pattern. Four qubit Dicke state was introduced in the context of studying superradiance. Mathematically, the four-qubit Dicke state is written in the following form
\begin{multline}
    |D^{(2)}_4\rangle = \frac{1}{\sqrt{6}} \left( |0011\rangle + |0101\rangle + |1001\rangle \right. \\
    \left. + |0110\rangle + |1010\rangle + |1100\rangle \right)
    \label{eq1}
\end{multline}  

Here, $|0\rangle$ and $|1\rangle$ represent the computational basis. The Dicke state is a widely used type of multi-partite entangled state in quantum information. However, very few works have been done on their nonlocality test. The test of nonlocality was carried out by generating the Dicke state by two distinct methods on the IBM QPUs. 
\subsection{Unitary Gate Based Initialization}
\begin{figure*}[t] 
    \centering
    \includegraphics[width=\linewidth]{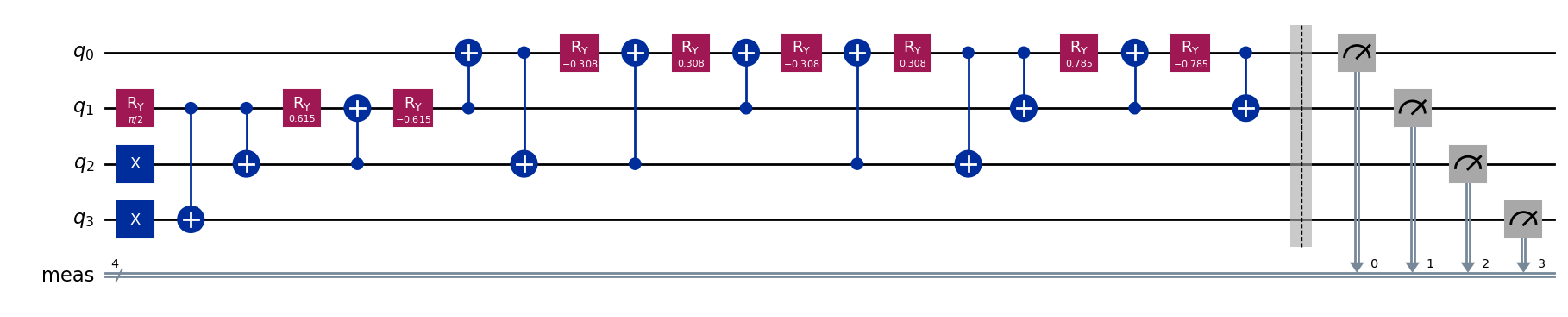}
    \caption{Illustration of a 4-qubit Dicke State ($|D^{(2)}_4\rangle$) state construction using gate based method.}
    \label{fig:dicke_state_gate_based}
\end{figure*}

\begin{figure*}[t] 
    \centering
    \includegraphics[width=\linewidth]{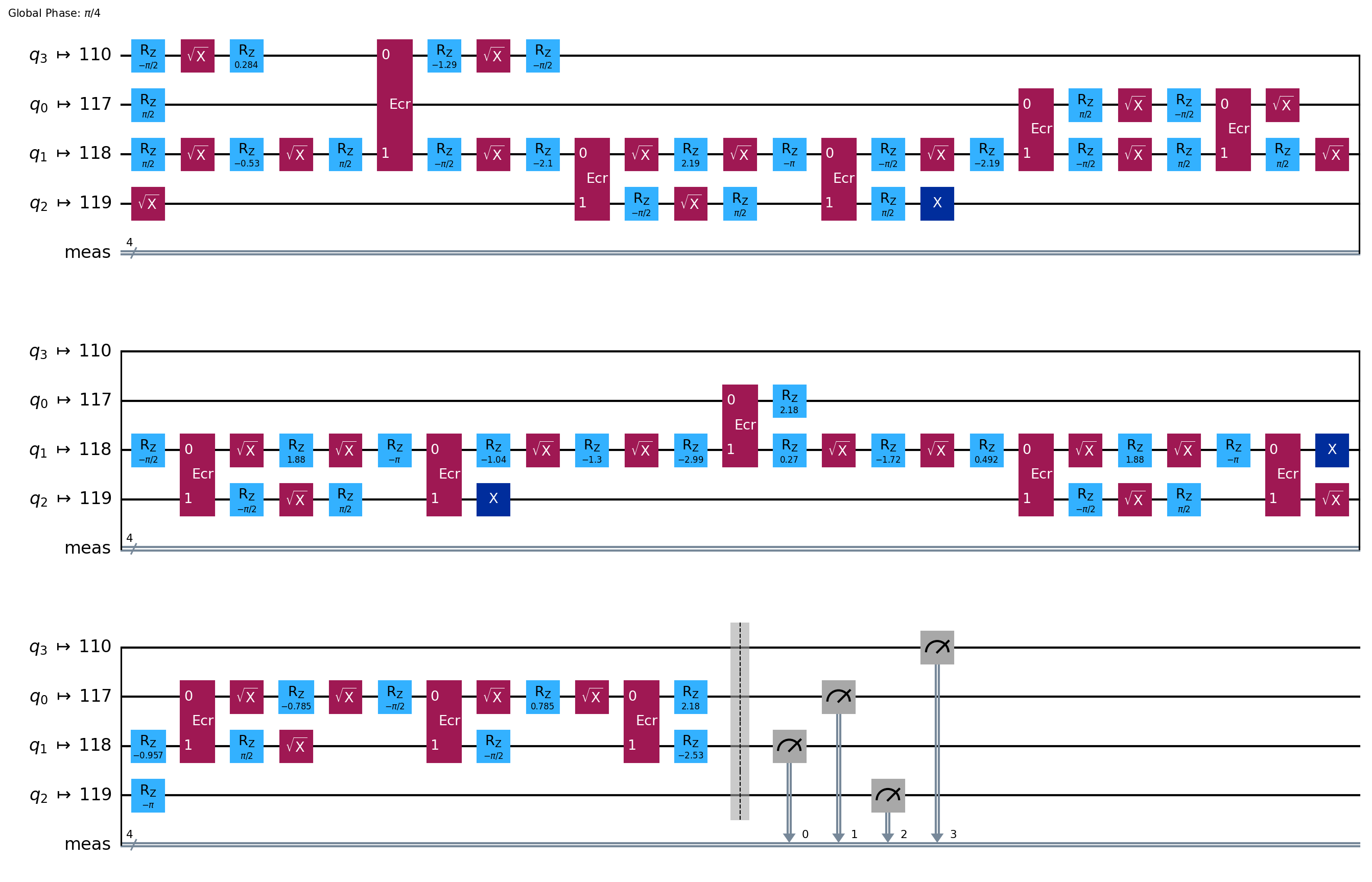}
    \caption{Illustration of Transpiled circuit (Unitary Gate Based prepared circuit) with optimization level 3 for \textit{ibm\_kyiv} QPU to find Expectation value of operator \texttt{XYXY} over 4 qubit Dicke state ($|D^{(2)}_4\rangle$).}
    \label{fig:dicke_state_transpiled_gate}
\end{figure*}

We generated the Bell state ($\vert\Phi^+$) using the Hadamard and CNOT gates. The same gate-based approach can be extended for the Dicke state as reported in \cite{9275336}. In this method, the construction of the Dicke state was implemented by a precise sequence of quantum gates, mainly via a combination of single-qubit rotations and controlled-NOT (CNOT) as shown in Fig. \ref{fig:dicke_state_gate_based}, and its transpile circuit with measurement of operator $XYXY$ is shown in Fig. \ref{fig:dicke_state_transpiled_gate}. 
We followed the approach described in \cite{9275336} for the precise generation of the four-qubit Dicke state as described in Eq. (\ref{eq1}), which closely resembles the theoretical Dicke state with minimal deviation. Obtained statevector using this method is

\begin{multline}
 |D^{(2)}_4\rangle =
    0.408269\,|0011\rangle + 0.408197\,|0101\rangle \\
    + 0.408211\,|1001\rangle + 0.408277\,|0110\rangle \\
    + 0.408291\,|1010\rangle + 0.408242\,|1100\rangle
    \label{statevector_using_gate_based_method}
\end{multline}

\subsection{State vector Based Initialization}
The generation of the Dicke state utilizes the framework established in \cite{PhysRevA.93.032318}. This method directly generates a circuit that prepares the specified Dicke state by constructing a state vector and deriving the associated quantum circuit that fulfills the generation of this vector. This technique considers the Dicke state as an isometry from a lower-dimensional Hilbert space, which may be effectively decomposed into a series of single-qubit and CNOT gates. This process works by zeroing out certain amplitudes using rotation and controlled gates. At each step, the vector is treated as if it were a single qubit's state, and the goal is to eliminate its contribution. Then the inverse of this set of gates is applied to zero states, and the desired state vector can be prepared. This method provides a more direct approach to achieving the desired state, without the necessity of fine-tuning gate parameters as necessary in \textbf{gate-based method}, and guarantees that the created quantum state is more precise. Obtained statevector using this method is
\begin{multline}
|D^{(2)}_4\rangle=
\frac{\sqrt{6}}{6} |0011\rangle+\frac{\sqrt{6}}{6} |0101\rangle+\frac{\sqrt{6}}{6} |0110\rangle \\
+\frac{\sqrt{6}}{6} |1001\rangle+\frac{\sqrt{6}}{6} |1010\rangle+\frac{\sqrt{6}}{6} |1100\rangle
\label{statevector_using_statevector_based_method}
\end{multline}

\begin{figure*} 
    \centering
    \includegraphics[width=0.85\linewidth]{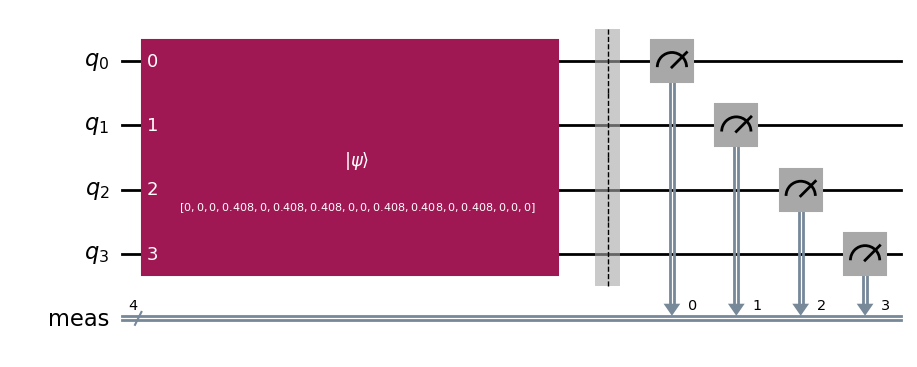}
    \caption{Illustration of a 4-qubit Dicke State ($|D^{(2)}_4\rangle$) state construction using state-vector based method.}
    \label{fig:dicke_state_qiskit_based}
\end{figure*}

\begin{figure*}[t] 
    \centering
    \includegraphics[width=\linewidth]{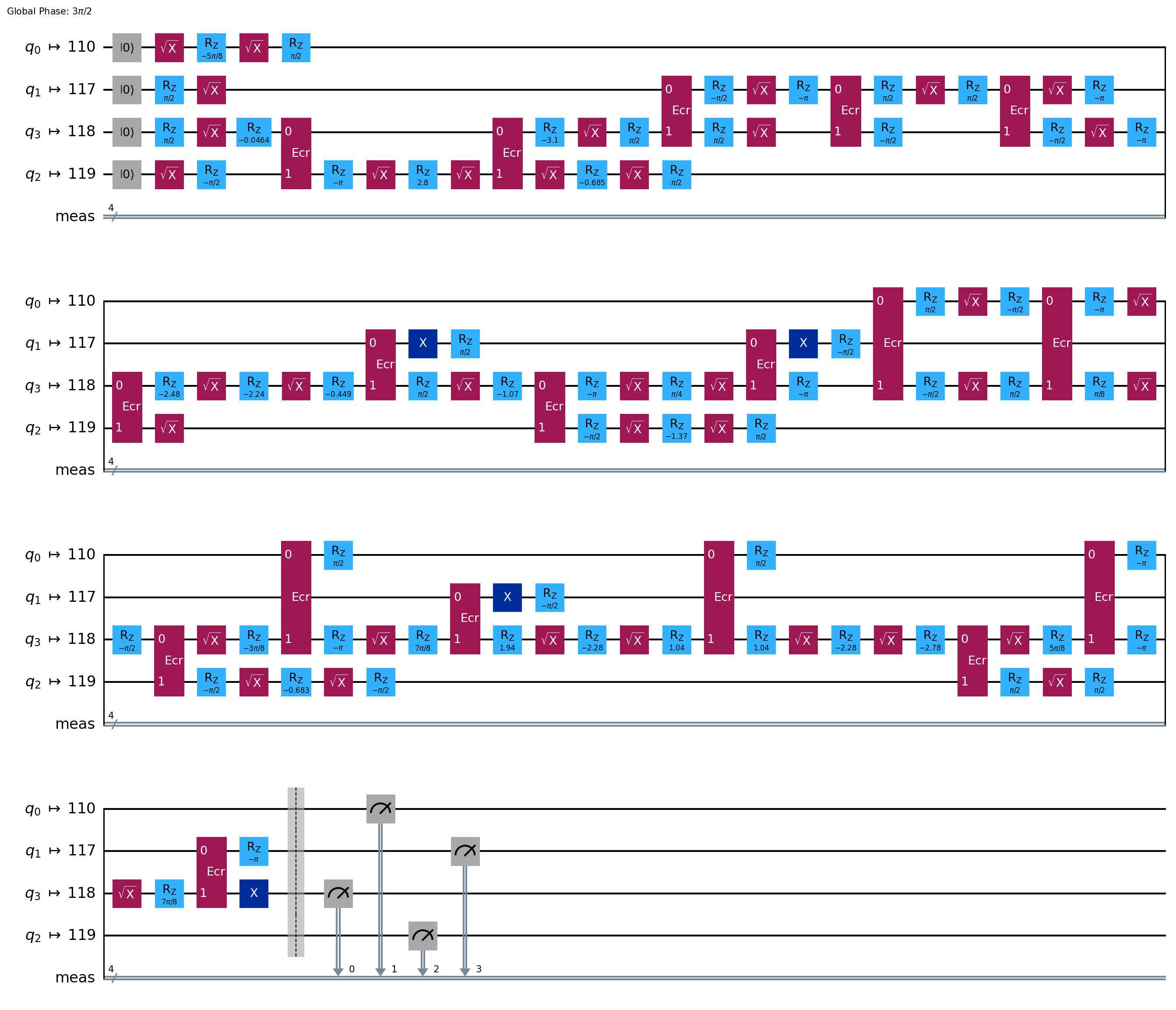}
    \caption{Illustration of Transpiled circuit (Statevector Based prepared circuit) with optimization level 3 for \textit{ibm\_kyiv} QPU to find Expectation value of operator \texttt{XYXY} over 4 qubit Dicke state ($|D^{(2)}_4\rangle$).}
    \label{fig:dicke_state_transpiled_state}
\end{figure*}

Fig. \ref{fig:dicke_state_qiskit_based} shows the generation of the Dicke state using this method, and Fig. \ref{fig:dicke_state_transpiled_state} shows the transpiled circuit for measuring the $XYXY$ operator.
Likewise, we have conducted the necessary measurements to violate the Bell inequality by altering the corresponding gates. The identical methodology is employed for circuits constructed via the state-vector-based method.

The method in \cite{PhysRevA.88.022126} devised the Bell inequality we construct. The inequality is given by
\begin{multline}
     |\hat{A} \hat{B} \hat{C} \hat{D} 
    + \hat{A} \hat{B}' \hat{C}' \hat{D}' \\
    + \hat{A}' \hat{B} \hat{C}' \hat{D} 
    - \hat{A}' \hat{B}' \hat{C} \hat{D}'| \leq 2.
    \label{dicke_bell}
\end{multline}

Where, The local observables $\hat{A}_i$, $\hat{B}_i$, $\hat{C}_i$, and $\hat{D}_i$ have the form: 
$\hat{X}_i = \mathbf{x}_i \cdot \boldsymbol{\sigma}$, where $\mathbf{x}_i = (x_{i1}, x_{i2}, x_{i3}) \in \mathbb{R}^3$ 
with $|\mathbf{x}_i| = 1$ (for $i = 0,1$), and the Pauli operator is given by 
$\boldsymbol{\sigma} = (\sigma_x, \sigma_y, \sigma_z)$.


\section{Development of Customized Operator}
To get the maximal violation of any Bell-type inequalities, we are assigning a local arbitrary observable in the form of a general operator $\boldsymbol{\hat{X}_{\hat{n}}}=x_i.\boldsymbol{\hat{\sigma}}$ where $x_i=(x_1,x_2,x_3)\in \mathbb{R}^3$ with $\sum_{i=1}^{3}x_{i}^2=1$ and Pauli operator $\boldsymbol{\hat{\sigma}}=(\hat{\boldsymbol\sigma_x},\hat{\boldsymbol\sigma_y},\hat{\boldsymbol\sigma_z})$. In a three-dimensional spherical space $\boldsymbol{\hat{X}_{\hat{n}}}$ can be represented as, 
\begin{multline}
    \boldsymbol{\hat{X}_{\hat{n}}} = \sin(\theta)\cos(\phi)\hat{\boldsymbol{\sigma}}_x 
    + \sin(\theta)\sin(\phi)\hat{\boldsymbol{\sigma}}_y + \cos(\theta)\hat{\boldsymbol{\sigma}}_z,
    \label{eq:oper}
\end{multline}
where $\phi\in(0,2\pi),\theta\in(0, \pi)$ are arbitrary angle spanning the three dimensional space. Using the above operator, we will show that the two-qubit Bell state and four-qubit Dicke state violate the Bell-type inequality both theoretically and experimentally on IBM QPUs.

\subsection{Maximum Violation Bound for two-Qubit \texorpdfstring{$\vert\Phi^+\rangle$}{Phi+} Bell State}


Here, we have derived the maximal violation condition for the maximally entangled two-qubit $\vert \Phi^+\rangle$ Bell state, which is written as,
\begin{equation}
\label{Eq2}
\vert\Phi^+\rangle=\frac{1}{\sqrt{2}}\vert00\rangle+\vert11\rangle,
\end{equation}
where  $|0\rangle$ and $|1\rangle$  represent the computational basis. The CHSH Bell inequality \cite{PhysRevLett.23.880} for two-particle has the following form,
\begin{equation}
\label{Eq3}
\vert S_{\vert\Phi^+\rangle}\vert=|\langle \hat{A}\hat{B}\rangle+\langle \hat{A}\hat{B'}\rangle-\langle \hat{A'}\hat{B}\rangle+\langle \hat{A'}\hat{B'}\rangle\vert\leq2,
\end{equation} 
where $\hat{A}(\hat{A'})$ and $\hat{B}(\hat{B'})$ are the arbitrary possible choices of measurements on particles 1 and 2, respectively. The verification procedure relies on computing all correlation functions, e.g., $ \langle \hat{A}\hat{B}\rangle$. The observable $\langle \hat{A}\hat{B}\rangle$ represents the expectation value of the joint probability of outcomes of measurements $\hat{A}$ and $\hat{B}$ on particle 1 and 2, respectively. Calculating all such terms appearing in Eq.(\ref{Eq3}) leads to the algebraic expression of the Bell polynomial. The value is then compared to the predicted value using local realism. In the domain of quantum mechanics, the measurements specified in Eq.(\ref{Eq3}) can be regarded as polarization measurements and are specified by the linear combination of Pauli operators in Eq.(\ref{eq:oper}). The quantum mechanical expectation value specified in $1^{st}$ term of Eq.(\ref{Eq3}) can be calculated for $\vert \Phi^+\rangle$ Bell state as,
\begin{align}
\label{Eq4}
\langle\hat{A}\hat{B}\rangle &= \cos(\theta_A) \cos(\theta_B) + \cos(\phi_A + \phi_B) \sin(\theta_A) \sin(\theta_B),
\end{align} 
where $\phi_i(\phi_{i}')$, $\theta_i(\theta_{i}')$, $\forall i \in(A,B)$ are polar angles specifying the measurement direction $\hat{n}_i,(\hat{n}_i')$. Similarly,  after calculating other terms in Eq.(\ref{Eq3}), we derived the Bell polynomial as shown in Eq.(\ref{Eq5}).

To get the values of polar angles corresponding to the maximal violation, we are required to maximize the Bell polynomial calculated in Eq.(\ref{Eq5}) using Mathematica\textsuperscript\textregistered. There is more than one optimum condition, out of which one condition is shown in Table (\ref{tab:T1}). By substituting the values of polar angles shown in Table (\ref{tab:T1})  in Eq.(\ref{Eq5}), we got the Bell value $\vert S_{\vert\Phi^+\rangle}\vert=2\sqrt{2}$. 



\begin{table}[h]
\centering
\begin{tabular}{|c|c|c|c|}
\hline
$\theta_A=45.03 ^{\circ}$ & $\theta_{A}'=44.98^{\circ}$ & $\theta_B=90.03^{\circ}$ & $\theta_{B}'=0.027^{\circ}$ \\
\hline
$\phi_A=0.014^{\circ}$ & $\phi_{A}'=180^{\circ}$ & $\phi_B=0.036^{\circ}$ & $\phi_{B}'=33.88^{\circ}$ \\
\hline
\end{tabular}
\caption{Value of optimum angles to get maximal violation for the two-qubit Bell state $|\Phi^+\rangle$.}
\label{tab:T1}
\end{table}

Using the given angle in Table \ref{tab:T1}, we calculated our customized operator in the form of Eq. \ref{eq:oper} and they are given below 
\begin{equation}
    \boldsymbol{\hat{A}} = 0.7074\hat{\boldsymbol{\sigma}}_x +0.0001 \hat{\boldsymbol{\sigma}}_y +0.7068 \hat{\boldsymbol{\sigma}}_z,
    \label{bell1}
\end{equation}
\begin{equation}
     \boldsymbol{\hat{B}} =1\hat{\boldsymbol{\sigma}}_x + 0.0006 \hat{\boldsymbol{\sigma}}_y -0.0004\hat{\boldsymbol{\sigma}}_z,
    \label{bell2}
\end{equation}

\begin{equation}
    \boldsymbol{\hat{A'}} = -0.7068\hat{\boldsymbol{\sigma}}_x + 0.0003 \hat{\boldsymbol{\sigma}}_y +0.7073\hat{\boldsymbol{\sigma}}_z,
    \label{bell3}
\end{equation}

\begin{equation}
    \boldsymbol{\hat{B'}} = 0.0004\hat{\boldsymbol{\sigma}}_x +0.0002 \hat{\boldsymbol{\sigma}}_y +1\hat{\boldsymbol{\sigma}}_z,
    \label{bell4}
\end{equation}

We implemented the above operators on two-qubit Bell states and executed them on the IBM QPU and showed Bell violation.

\subsection{Maximum violation bound for Four-qubit Dicke state }
By extending our approach to higher qubits, we took a four-qubit Dicke state ($|D^{(2)}_4\rangle$)  in this section to get the violation of the four-qubit Bell inequality. The four qubit  Dicke state ($|D^{(2)}_4\rangle$) is defined in Eq. \ref{eq1}. The Bell-type inequality for the $|D^{(2)}_4\rangle$ state is defined in Eq.\ref{dicke_bell},  where $A(A')$,$B(B')$,$C(C')$ and $D(D')$ are the arbitrary possible choice of measurements on particle 1, 2, 3, and 4 respectively.
Similarly, we calculated quantum mechanical expectation values of all terms given in Eq.(\ref{dicke_bell}) to find the value of Bell-Polynomial for $|D^{(2)}_4\rangle$ state. We calculated the quantum mechanical expectation value of the  $1^{st}$ term of Eq.(\ref{dicke_bell}), and the same is given in Eq. (\ref{b2}).

Similarly, we calculated all four terms present in Eq.(\ref{dicke_bell}). After the simplification, by employing  Mathematica\textsuperscript\textregistered for finding the polar angles corresponding to the maximum violation of the four-qubit Dicke state. Similarly, we calculated the polar angles in which maximum violation for the qubits Dicke state occurred as we did for the two-qubit state and shown in the Table (\ref{t2}) below. Theoretically, the maximum violation for the Dicke state is \textbf{3.055}. From the Polar angles calculated from the above equation, we construct the eight customized operators for violating the Bell inequality in IBM QPUs. 
\begin{table}[h]
\centering
\begin{tabular}{|c|c|c|c|}
\hline
$\theta_A=107.792 ^{\circ}$ & $\theta_{A}'=30.3962^{\circ}$ & $\theta_B=107.793^{\circ}$ & $\theta_{B}'=69.0948^{\circ}$ \\
$\theta_C=30.3962 ^{\circ}$ & $\theta_{C}'=107.792^{\circ}$ & $\theta_D=107.793^{\circ}$ & $\theta_{D}'=69.0964^{\circ}$\\
\hline
$\phi_A=57.2234^{\circ}$ & $\phi_{A}'=57.2234^{\circ}$ & $\phi_B=57.2234^{\circ}$ & $\phi_{B}'=57.2234^{\circ}$\\
$\phi_C=57.2234^{\circ}$ & $\phi_{C}'=57.2234^{\circ}$ & $\phi_D=57.2234^{\circ}$ & $\phi_{D}'=57.2234^{\circ}$\\
\hline
\end{tabular}
\caption{Value of optimum angles to get maximal violation for the four-qubit Dicke state $|D^{(2)}_4\rangle$.}
\label{t2}
\end{table}


These eight operators are given below,
\begin{equation}
    \boldsymbol{\hat{A}} = 0.515473 \hat{\boldsymbol{\sigma}}_x + 0.800575\hat{\boldsymbol{\sigma}}_y - 0.305561 \hat{\boldsymbol{\sigma}}_z,
    \label{d13}
\end{equation}
\begin{equation}
    \boldsymbol{\hat{B}} = 0.515469 \hat{\boldsymbol{\sigma}}_x +0.800568\hat{\boldsymbol{\sigma}}_y -0.305585 \hat{\boldsymbol{\sigma}}_z,
\end{equation}
\begin{equation}
    \boldsymbol{\hat{C}} = 0.273919 \hat{\boldsymbol{\sigma}}_x +0.425419\hat{\boldsymbol{\sigma}}_y+0.862547 \hat{\boldsymbol{\sigma}}_z,
\end{equation}
\begin{equation}
    \boldsymbol{\hat{D}} =  0.515469 \hat{\boldsymbol{\sigma}}_x + 0.800568\hat{\boldsymbol{\sigma}}_y-0.305585 \hat{\boldsymbol{\sigma}}_z,,
\end{equation}
\begin{equation}
    \boldsymbol{\hat{A'}} = 0.273918 \hat{\boldsymbol{\sigma}}_x +0.425419\hat{\boldsymbol{\sigma}}_y+0.862547 \hat{\boldsymbol{\sigma}}_z,
\end{equation}
\begin{equation}
    \boldsymbol{\hat{B'}} = 0.505728 \hat{\boldsymbol{\sigma}}_x +0.78544\hat{\boldsymbol{\sigma}}_y+0.356822 \hat{\boldsymbol{\sigma}}_z,
\end{equation}
\begin{equation}
    \boldsymbol{\hat{C'}} = 0.515473 \hat{\boldsymbol{\sigma}}_x +0.800575\hat{\boldsymbol{\sigma}}_y-0.305561 \hat{\boldsymbol{\sigma}}_z,
\end{equation}
\begin{equation}
    \boldsymbol{\hat{D'}} = 0.505734 \hat{\boldsymbol{\sigma}}_x +0.785449\hat{\boldsymbol{\sigma}}_y+0.356796 \hat{\boldsymbol{\sigma}}_z,
    \label{d20}
\end{equation}

\section{Experimental Implementation}
In this section, we discuss the implementation of our customized operator for violation of the Bell-type inequalities for the two-qubit Bell state, as well as the four-qubit Dicke state. Firstly, we discuss the two qubits, and then we extend our approach to the four qubits state.
\subsection{Two-qubit Bell state implementation}
A two-qubit Bell inequality experiment can be performed using different measurement settings. These measurement settings are typically represented by the operators $A$, $B$, $A'$, and $B'$ acting on the first and second qubits, respectively. To evaluate quantum inequalities such as those tested in this experiment, we decompose the operators $A$, $B$, $A'$, and $B'$ in terms of the identity matrix and the Pauli operators. Each operator is expressed as given in Eq. (\ref{eq:oper}).
Where, $\sigma_x$, $\sigma_y$, and $\sigma_z$ are the Pauli matrices. This decomposition allows us to express each operator as a linear combination of fundamental quantum operations, enabling a more systematic way to compute expectation values and test for inequality violations. The expression of these operators is given in Eq. (\ref{bell1}-\ref{bell4}).

Since each operator is decomposed in terms of three Pauli matrices, the tensor product of two operators (e.g., $  A\otimes B$) results in a total of 9 unique operators. 
\[
k_i k_j = k_i \otimes k_j, \quad \text{where } k_i, k_j \in \{\sigma_x, \sigma_y, \sigma_z\}
\]
Such as $\sigma_x \otimes \sigma_x$, $\sigma_x \otimes \sigma_y$, ... , $\sigma_z \otimes \sigma_z$.  Each operator corresponds to a distinct combination of measurements on the two qubits, enabling the calculation of expectation values for every pair of measurement configurations. However, not all these operators contribute equally to the quantum state under study, so we need to identify those that yield non-zero expectation values. To make this process more efficient, we compute the expectation value of each operator when acting on the two-qubit Bell state $|\Phi^+\rangle$. Through this analysis, we find that only three operators $XX, YY, ZZ$ have non-zero expectation values. These operators are, therefore, sufficient for our analysis and significantly reduce the computational overhead.

\subsubsection{Qiskit Implementation}

After identifying the relevant operators, the implementation of the inequality violation test was carried out using Qiskit, which is a comprehensive open-source quantum computing framework\cite{qiskit2024}. We implemented the experiment using the \texttt{SamplerV2} primitive in Qiskit’s IBM Quantum runtime. The quantum circuit is constructed to prepare the desired entangled Bell state, followed by the application of the decomposed operators. We use Qiskit’s \texttt{transpiler} to optimize the circuit with optimization level 3, minimizing the circuit depth (number of sequential gate layers, where parallelizable operations are grouped) and reducing the number of two-qubit gates.

The optimized circuits are then executed on the IBM QPU, with each circuit being sampled 10,000 times to gather sufficient statistics for the calculation of expectation values. This high shot count ensures that the computed expectation values are robust against statistical fluctuations. To complement the real hardware, the experiments were also simulated on noisy simulators, which replicate the features and noise characteristics of the actual IBM QPU \cite{PhysRevA.89.042123}. The noisy simulator enabled multiple runs without hardware access limitations.

\subsubsection{Expectation Value Calculation from Measurement Counts}

Upon execution, Qiskit’s \texttt{Sampler} returns a dictionary of raw measurement counts \(\{\text{bitstring}: \text{counts}\}\). Denoting the count of each bitstring \(b\) by \(\text{counts}(b)\) and the total shot count by \(N\), we compute the empirical probability
\[
P(b) = \frac{\text{counts}(b)}{N}\,.
\]
Each bitstring \(b\) is then assigned an eigenvalue \(\lambda_O(b)\in\{+1,-1\}\) according to the Pauli‑decomposition of the operator \(O\). The expectation value is obtained as the weighted sum
\[
\langle O \rangle = \sum_{b} P(b)\,\lambda_O(b)\,.
\]
More details are provided in Appendix C.

\subsubsection{Correcting for Noise and Mitigating Errors}

Quantum circuits executed on superconducting qubit devices suffer primarily from two types of errors. Gate errors arise due to Imperfections in control pulses, crosstalk between qubits, and decoherence during gate operations, which introduce both coherent and stochastic noise into the circuit. Another type of error is measurement error, which is the result of Readout infidelity in the dispersive measurement process, can misclassify computational basis states, leading to bit‑flip and assignment errors.

To improve the accuracy of measured expectation values, we apply Dynamic Decoupling (DD) and Twirling for gate‑error and Matrix‑free measurement mitigation (MThree) for measurement‑error mitigation.

Dynamic decoupling inserts sequences of carefully timed $\pi$‑pulses (e.g., CPMG or XY sequences) between computational gates\cite{PhysRevApplied.20.064027}. These pulses periodically refocus qubit phases to average out low‑frequency environmental noise, prolonging coherence during long or complex gate sequences and reducing errors from slow drift and static biases.

Randomized twirling combats coherent gate errors by surrounding each target gate with randomly chosen Pauli operators\cite{PhysRevA.94.052325}. By averaging over many such random instances, coherent over‑ and under‑rotations are converted into stochastic Pauli noise channels. This symmetrization simplifies the error model to a depolarizing channel, making error characteristics easier to estimate and correct via Pauli‑frame updates.

We employ matrix‑free measurement mitigation (MThree) \cite{PRXQuantum.2.040326} to correct for readout errors. MThree builds a calibration matrix by preparing and measuring all computational basis states to estimate confusion probabilities. Rather than explicitly inverting this matrix, its iterative, matrix‑free algorithm applies correction filters directly to the raw counts, yielding quasi‑probabilities that more faithfully represent the true measurement statistics.

While gate‑error mitigation methods such as Dynamic Decoupling and Randomized Twirling are applied at circuit‑execution time and require no further post-processing, measurement‑error mitigation using MThree does involve post‑processing. First, the backend is calibrated by preparing and measuring all computational basis states to estimate the measurement confusion matrix. These calibration results are then used to adjust the raw counts: the matrix‑free MThree algorithm applies correction filters to yield quasi‑probabilities, which are used to recompute the counts. This produces a more reliable estimate of the true measurement statistics.

\subsection{Four qubit Dicke state implementation}
We extended our approach to a four-qubit state. In the four-qubit case, the decomposition of operators follows the same general principle as in the two-qubit case. Each operator denoted as $A$, $B$, $C$, $D$, $A'$, $B'$, $C'$ and $D'$ is decomposed into a combination of Pauli matrices and the identity, according to Eq.(\ref{eq:oper}). The main difference lies in the increased complexity due to the larger system size and the nature of the quantum state involved. Specifically, we now investigated with a four-qubit Dicke state as described in Eq.(\ref{eq1}) instead of a Bell state.
In this case, we compute the tensor product of four sets of decomposed operators. These operators for the four-qubit Dicke state are defined in Eq. (\ref{d13}-\ref{d20}).
\begin{equation}
\begin{split}
k_i k_j k_l k_m = k_i \otimes k_j \otimes k_l \otimes k_m, \quad \\
\text{where } k_i, k_j, k_l, k_m \in \{ \sigma_x, \sigma_y, \sigma_z\}
\end{split}
\end{equation} 

This results in a total of 81 unique operator combinations, such as $\sigma_x\otimes \sigma_x \otimes \sigma_x \otimes \sigma_x$, $\sigma_x\otimes \sigma_x \otimes \sigma_x \otimes \sigma_y$, ... , $\sigma_z\otimes \sigma_z \otimes \sigma_z \otimes \sigma_z$. However, based on the state being studied, certain operator combinations will yield zero expectation values, and we focus only on those that contribute to the final expectation value calculation. For the four-qubit Dicke state, we found 21 such operators out of the 81 operators.
The implementation on Qiskit follows the same procedure as described for the two-qubit case but is scaled up to accommodate the higher qubit (four-qubit Dicke state) count.

\section{Results and Discussion}
\subsubsection{Two Qubit Bell state}
\begin{figure}[h]
    \centering
    \includegraphics[width=1\linewidth]{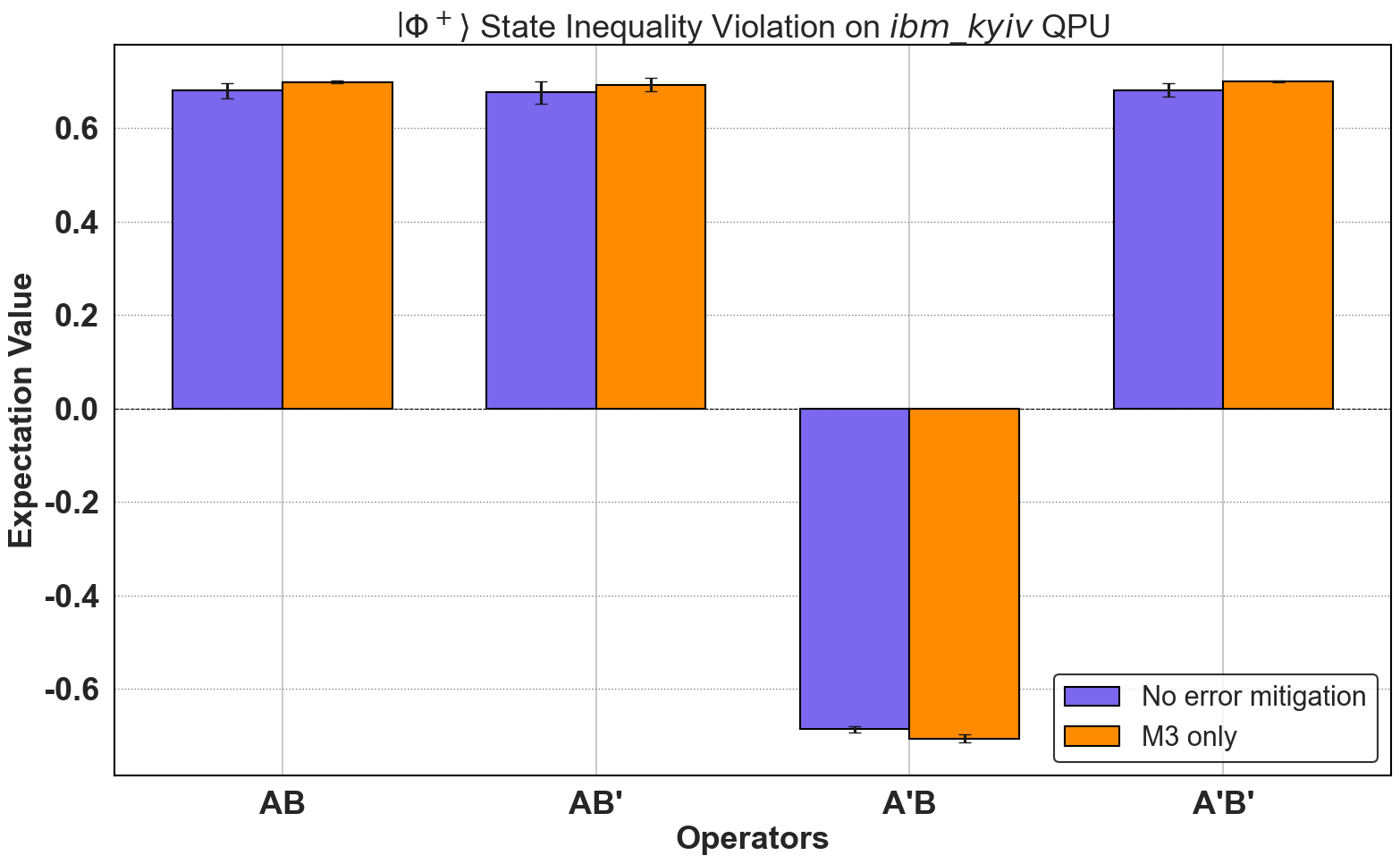}
    \caption{A comparison of the average expectation values with and without error mitigation of all four operators for CHSH violation for the two-qubit Bell state on ibm\_kyiv QPU. }
    \label{bell_op_kyiv}
\end{figure}

\begin{figure}[h]
    \centering
    \includegraphics[width=1\linewidth]{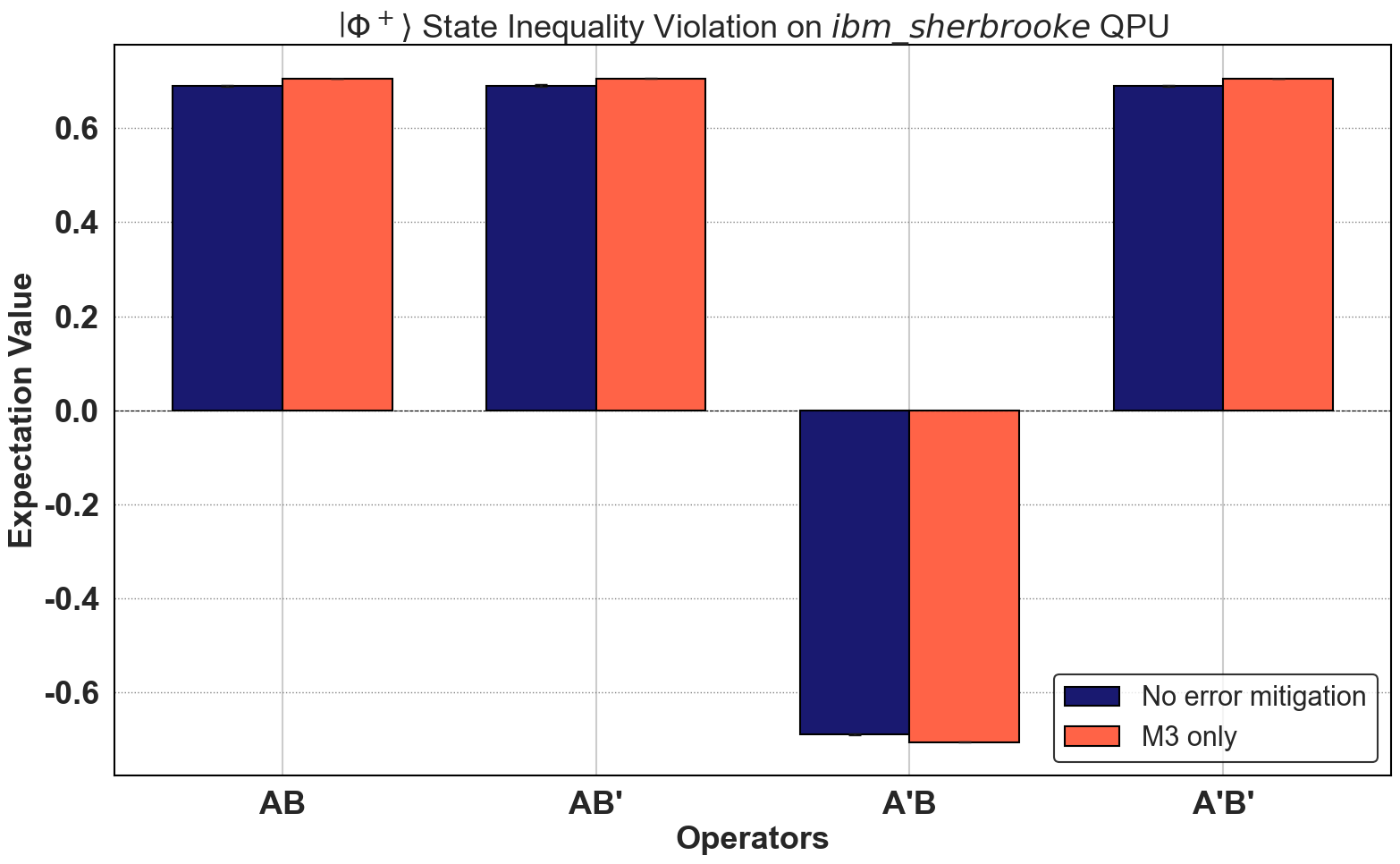}
    \caption{A comparison of the average expectation values with and without error mitigation of all four operators for CHSH violation for the two-qubit Bell state on ibm\_sherbrook QPU. }
    \label{bell_op_sherbrook}
\end{figure}

\begin{figure}[]
    \centering
    \includegraphics[width=1\linewidth]{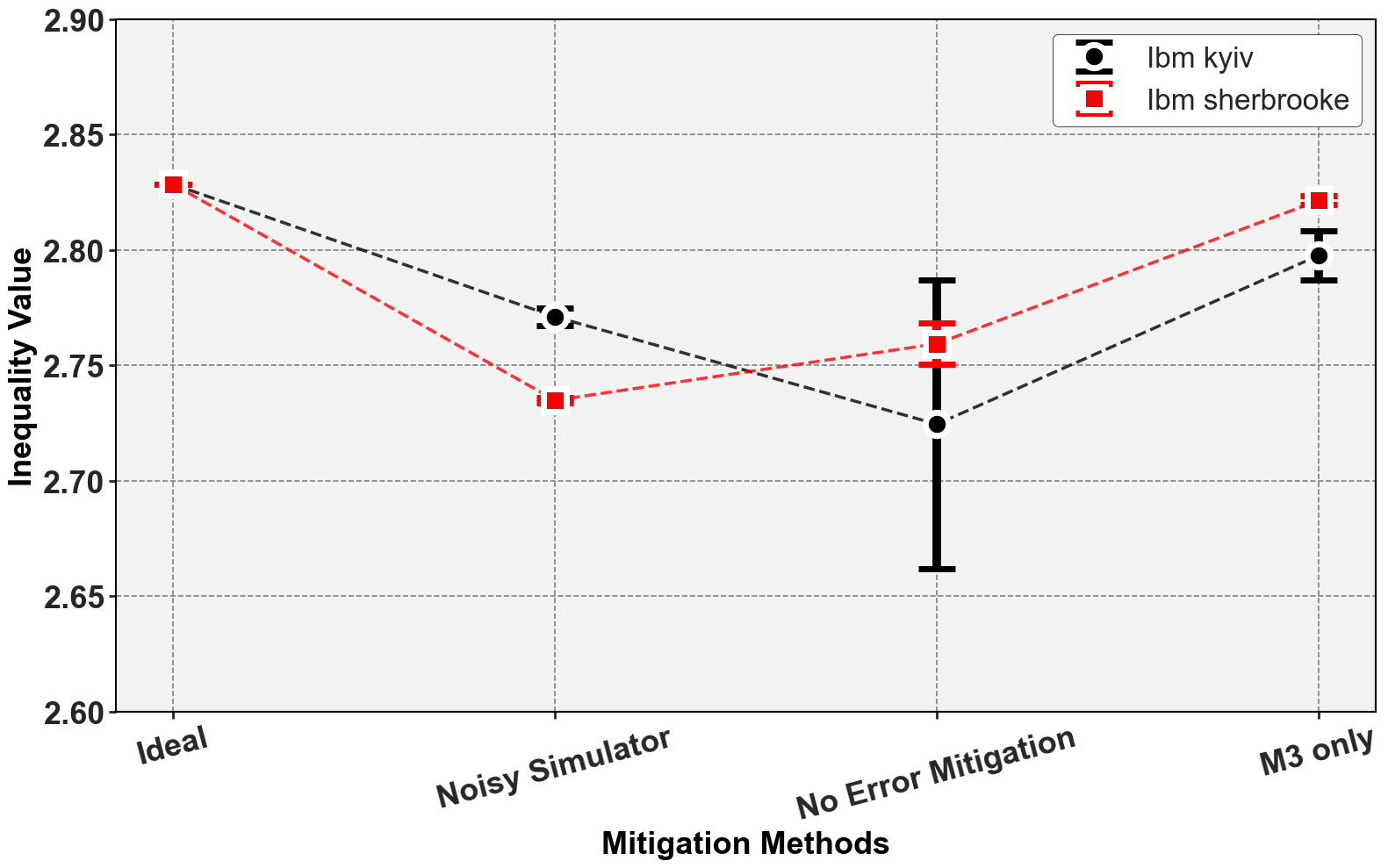}
    \caption{A comparison of the CHSH-Bell values of Ideal Simulator, noisy, and real hardware with and without mitigation methods on two IBM QPUs.}
    \label{bell_comp}
\end{figure}

We employed our customized operator to evaluate the violation of the Bell inequality for the two-qubit Bell state. We generate the two-qubit Bell state $\vert\Phi^+\rangle$ using Hadamard and CNOT gates as defined in Eq. (\ref{Eq2}). We applied our customized operators $A$, $A'$, $B$, and $B'$ in the respective first and second qubits as discussed in the previous section. The CHSH Bell inequality is defined in Eq. (\ref{Eq3}) corresponding to the four measurement operators $AB$, $AB'$, $A'B$, and $A'B'$. For the violation of the CHSH Bell inequality using our customized operator, we use only three operators defined in the previous section; we ran experiments on two IBM quantum computers: \texttt{ibm\_kyiv} and \texttt{ibm\_sherbrook}. We also used their simulated counterparts known as noisy simulator, by applying the noise model of \texttt{ibm\_kyiv} and \texttt{ibm\_sherbrooke} to the simulator, which let us model the effects of noise without actually running on real hardware. On top of that, we compared all these results to ideal outputs from a noise-free simulator, giving us a clear benchmark for how the real devices and their simulated versions perform.
For each setup---ideal, simulated (noisy), and real hardware---we ran the circuit three times on different days and times for the robust violation, using 10,000 shots per run. On the real quantum devices, we also tested the effect of measurement error mitigation using the M3 method, so we could see how much it helps improve the results. We executed the circuit after calibration of QPUs to analyze the calibration data, which is given in the appendix in Tables \ref{tab:calibration_kyiv} and \ref{tab:calibration_sherbrooke}.
The average expectation values from all these runs are shown in Tables~\ref{tab:kyiv_results_bell} and \ref{tab:sherbrook_results_bell}, covering the  ibm\_kyiv and ibm\_sherbrooke devices. Fig. \ref{bell_op_kyiv} and \ref{bell_op_sherbrook} show the comparison of the expectation values of four operators, with and without mitigation, on real IBM QPUs, ibm\_kyiv and ibm\_sherbrook, respectively. In our investigation, ibm\_sherbrook is giving better results compared to the ibm\_kyiv for the two-qubit Bell state, without and with M3 mitigation. This is due to different noise levels at the time of execution. Using M3 error mitigation on \texttt{ibm\_sherbrook}, the observed CHSH inequality violation reached a value of $2.821 \pm 0.010$, which lies within $0.7\sigma$ of the theoretical maximum value of $2.828$, indicating a strong agreement with the expected quantum prediction.  Fig.~\ref{bell_comp} compares the CHSH results, 
 for Ideal, Noisy, without and with M3 mitigation technique from the two IBM QPUs.
We attempted to observe the violation of the CHSH inequality using all nine operator combinations ($XX$, $XY$, ..., $ZZ$). However, no violation was observed. Our investigation suggests that when all nine operators are employed, the operator that has zero expectation values contributes and leads to the nonviolation. 

\begin{table*}[ht]
\centering
\begin{tabular}{|l|c|c|c|c|}
\hline
\textbf{Operator} & \textbf{Ideal} & \multicolumn{3}{c|}{\textbf{Two qubit - IBM\_Kyiv}} \\
\cline{3-5}
                  &                & \textbf{Simulator (Noisy)} & \textbf{Without Mitigation} & \textbf{M3} \\
\hline
AB       & 0.707  & 0.692 $\pm$ 0.000 & 0.680 $\pm$ 0.016 & 0.698 $\pm$ 0.003 \\
AB$'$    & 0.707  & 0.692 $\pm$ 0.001 & 0.676 $\pm$ 0.024 & 0.693 $\pm$ 0.013 \\
A$'$B    & -0.707 & -0.695 $\pm$ 0.001 & -0.685 $\pm$ 0.007 & -0.705 $\pm$ 0.008 \\
A$'$B$'$ & 0.707  & 0.692 $\pm$ 0.000 & 0.681 $\pm$ 0.015 & 0.699 $\pm$ 0.001 \\
\hline
\textbf{Inequality} & 2.828 & 2.771 $\pm$ 0.003 & 2.724 $\pm$ 0.062 & 2.790 $\pm$ 0.010 \\
\hline
\end{tabular}
\caption{Two-qubit CHSH test results on IBM\_Kyiv with and without M3 mitigation}
\label{tab:kyiv_results_bell}
\end{table*}

\begin{table*}[ht]
\centering
\begin{tabular}{|l|c|c|c|c|}
\hline
\textbf{Operator} & \textbf{Ideal} & \multicolumn{3}{c|}{\textbf{Two qubit - IBM\_Sherbrook}} \\
\cline{3-5}
                  &                & \textbf{Simulator (Noisy)} & \textbf{Without Mitigation} & \textbf{M3} \\
\hline
AB       & 0.707  & 0.685 $\pm$ 0.000 & 0.689 $\pm$ 0.002 & 0.705 $\pm$ 0.0003 \\
AB$'$    & 0.707  & 0.686 $\pm$ 0.000 & 0.690 $\pm$ 0.002 & 0.705 $\pm$ 0.0001 \\
A$'$B    & -0.707 & -0.659 $\pm$ 0.000 & -0.689 $\pm$ 0.001 & -0.704 $\pm$ 0.0007 \\
A$'$B$'$ & 0.707  & 0.703 $\pm$ 0.000 & 0.689 $\pm$ 0.002 & 0.705 $\pm$ 0.0005 \\
\hline
\textbf{Inequality} & 2.828 & 2.734 $\pm$ 0.001 & 2.759 $\pm$ 0.008 & 2.821 $\pm$ 0.0019 \\
\hline
\end{tabular}
\caption{Two-qubit CHSH test results on IBM\_Sherbrook with and without M3 mitigation}
\label{tab:sherbrook_results_bell}
\end{table*}


\subsubsection{Four qubit Dicke state}
We then focused on estimating the Bell-type inequality violation for the four-qubit Dicke state, as defined in Eq.~(\ref{eq1}). This state was prepared using two different approaches: a unitary gate-based method and a statevector-based method, which were discussed in the previous section. To assess the presence of non-classical correlations, we used the Bell-type inequality introduced in Eq.~(\ref{dicke_bell}). We measured the expectation values of four operator combinations: \(ABCD\), \(AB'C'D'\), \(A'BC'D\), and \(A'B'CD'\) defined in Eq. (\ref{dicke_bell}). Using our customized operators discussed in the previous section we found the theoretical maximum values of these operators are 0.800, 0.930, 0.800, and -0.525, respectively. Summing these expectation values yields a Bell parameter of $3.055$, representing the theoretical maximum violation of the Bell-type inequality achievable by the Dicke state.

\begin{figure}
    \centering
    \includegraphics[width=1\linewidth]{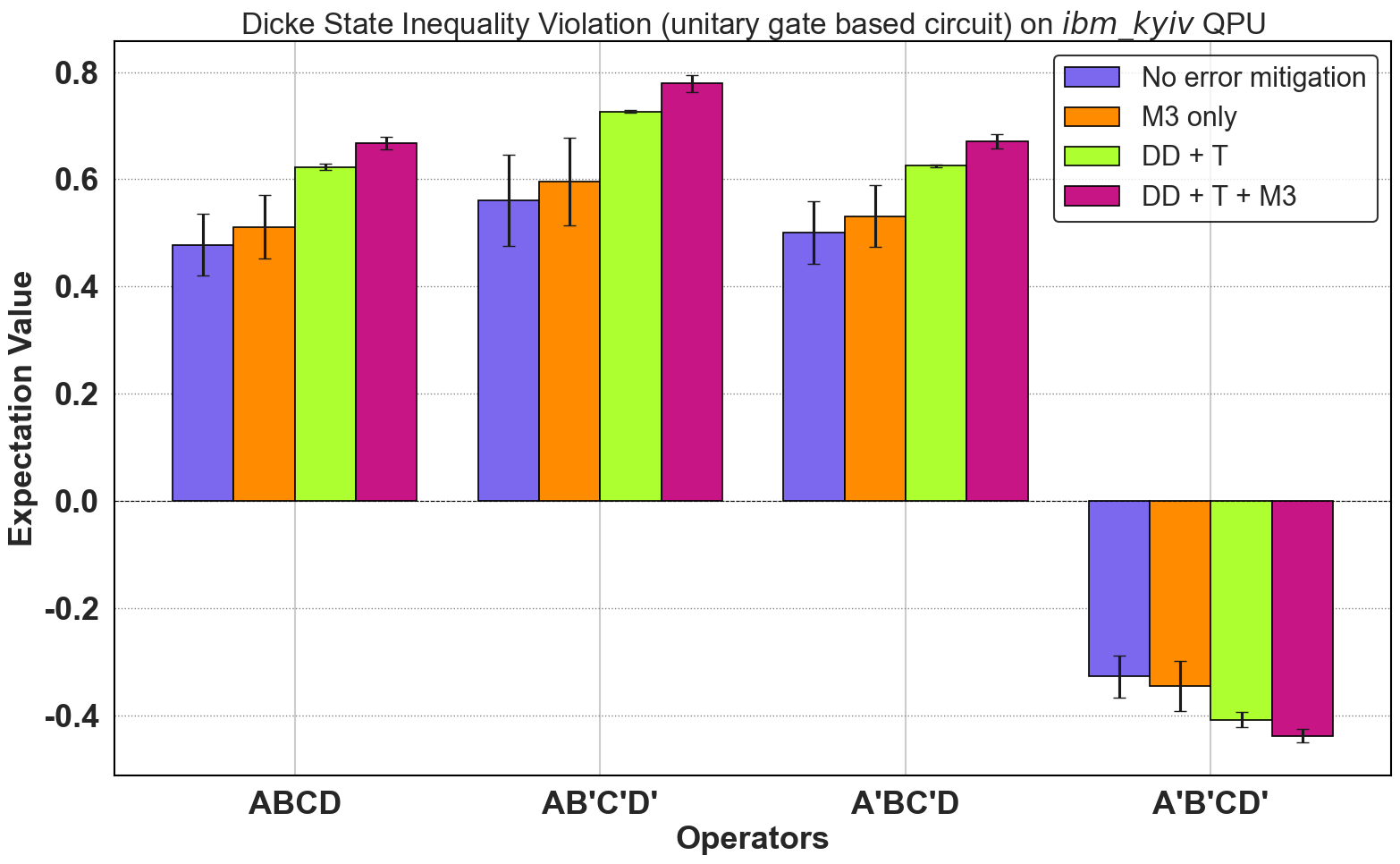}
    \caption{A comparison of the average expectation values of the measurement operator for the violation of the Bell-type inequality without and with various mitigation techniques on ibm\_kyiv, using the gate-based method.}
    \label{gate_four_op_Dicke_kyiv}
\end{figure}

\begin{figure}
    \centering
    \includegraphics[width=1\linewidth]{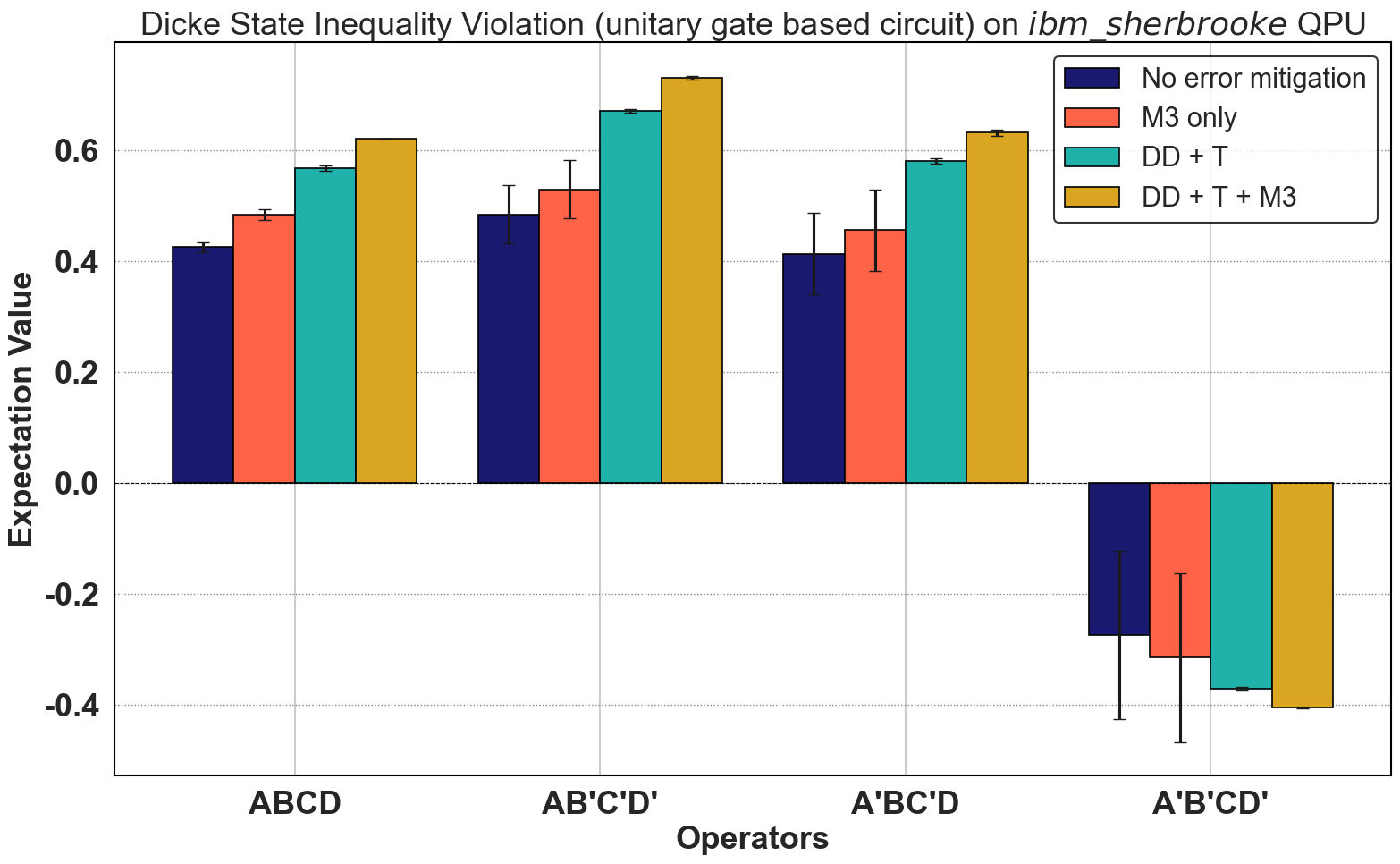}
    \caption{A comparison of the average expectation values of the measurement operator for the violation of the Bell-type inequality without and with various mitigation techniques on ibm\_sherbrook, using the gate-based method.}
    \label{gate_four_op_Dicke_sherbrook}
\end{figure}

For the Dicke state prepared by the unitary gate-based method, we evaluated the inequality using the same two IBM QPUs, \texttt{ibm\_kyiv} and \texttt{ibm\_sherbrooke}. We used only 21 operators for finding the expectation values of all the measurement operators defined in Eq. \ref{dicke_bell}. Each experimental setup—including the ideal (noise-free) simulator, noisy simulator, and real IBM QPUs—was executed three times with 10,000 shots per run on different days and times to verify robust violations. Each time, we took the calibration data of the QPUs, and the same is given in the appendix. We evaluated the results both without any error mitigation and with various mitigation techniques such as dynamical decoupling, twirling, and M3 (measurement error mitigation), as described in earlier sections.
Tables~\ref{tab:gate_dicek_sherbrook} and \ref{gate_dicke_Kyiv} present the average expectation values of all four measurement settings, as well as the resulting Bell parameters for both quantum devices. From Tables~\ref{tab:gate_dicek_sherbrook} and \ref{gate_dicke_Kyiv}, it is evident that error mitigation plays a crucial role in recovering the Bell parameter. Without any mitigation techniques, a violation of the Bell-type inequality was not observed on either of the IBM QPUs. However, \texttt{ibm\_kyiv} produced a value closer to the classical bound of 2. As we applied more advanced error mitigation methods, the results improved linearly on both devices. This is mainly because these methods help reduce errors from the gates and the hardware itself. Fig.~\ref{gate_four_op_Dicke_kyiv} and \ref{gate_four_op_Dicke_sherbrook} present a comparison of the average expectation values for different error mitigation techniques applied to the measurement operators on \texttt{ibm\_kyiv} and \texttt{ibm\_sherbrook}, respectively.

\begin{table*}[]

\begin{tabular}{|l|c|c|c|c|c|c|}
\hline
\textbf{Operator} & \textbf{Ideal} & \multicolumn{5}{c|}{\textbf{Gate-Based IBM\_Sherbrook}} \\
\cline{3-7}
                  &                & \textbf{Simulator (Noisy)} & \textbf{Without Mitigation} & \textbf{M3} & \textbf{DD+T} & \textbf{DD+T+M3} \\

\hline
ABCD              & 0.800   & 0.668 ± 0.004   & 0.425 ± 0.009   & 0.483 ± 0.009   & 0.567 ± 0.004   & 0.621 ± 0.0006 \\
AB$'$C$'$D$'$     & 0.930   & 0.776 ± 0.006   & 0.484 ± 0.052   & 0.529 ± 0.052   & 0.670 ± 0.002   & 0.731 ± 0.003 \\
A$'$BC$'$D        & 0.800   & 0.668 ± 0.003   & 0.413 ± 0.073   & 0.456 ± 0.073   & 0.580 ± 0.004   & 0.632 ± 0.005 \\
A$'$B$'$CD$'$     & -0.525  & -0.441 ± 0.003  & -0.274 ± 0.152  & -0.315 ± 0.152  & -0.371 ± 0.003  & -0.405 ± 0.0008 \\
\hline
\textbf{Inequality} & 3.055 & 2.554 ± 0.012   & 1.598 ± 0.177   & 1.785 ± 0.177   & 2.190 ± 0.005   & 2.390 ± 0.008 \\
\hline
\end{tabular}
\caption{Gate-based method with all results in IBM Sherbrooke using 21 operators}
\label{tab:gate_dicek_sherbrook}
\end{table*}

\begin{table*}[]

\begin{tabular}{|l|c|c|c|c|c|c|}
\hline
\textbf{Operator} & \textbf{Ideal} & \multicolumn{5}{c|}{\textbf{Gate-Based IBM\_Kyiv}} \\
\cline{3-7}
                  &                & \textbf{Simulator (Noisy)} & \textbf{Without Mitigation} & \textbf{M3} & \textbf{DD+T} & \textbf{DD+T+M3} \\

\hline
ABCD              & 0.800   & 0.673 ± 0.002   & 0.477 ± 0.057   & 0.511 ± 0.059   & 0.622 ± 0.005   & 0.667 ± 0.012 \\
AB$'$C$'$D$'$     & 0.930   & 0.785 ± 0.001   & 0.561 ± 0.085   & 0.594 ± 0.081   & 0.726 ± 0.002   & 0.778 ± 0.016 \\
A$'$BC$'$D        & 0.800   & 0.674 ± 0.001   & 0.501 ± 0.058   & 0.531 ± 0.056   & 0.625 ± 0.006   & 0.670 ± 0.013 \\
A$'$B$'$CD$'$     & -0.525  & -0.438 ± 0.003  & -0.327 ± 0.039  & -0.345 ± 0.047  & -0.408 ± 0.014  & -0.438 ± 0.011 \\
\hline
\textbf{Inequality} & 3.055 & 2.571 ± 0.005   & 1.862 ± 0.197  & 1.982 ± 0.195   & 2.382 ± 0.019   & 2.555 ± 0.049 \\
\hline
\end{tabular}

\caption{Gate-based method with all results in IBM Kyiv using 21 operators}
\label{gate_dicke_Kyiv}
\end{table*}

\begin{table*}[]

\begin{tabular}{|l|c|c|c|c|c|c|}
\hline
\textbf{Operator} & \textbf{Ideal} & \multicolumn{5}{c|}{\textbf{Statevector IBM\_Kyiv}} \\
\cline{3-7}
                  &                & \textbf{Simulator (Noisy)} & \textbf{Without Mitigation} & \textbf{M3} & \textbf{DD+T} & \textbf{DD+T+M3} \\

\hline
ABCD         & 0.800  & 0.688 ± 0.000   & 0.654 ± 0.010   & 0.707 ± 0.005   & 0.626 ± 0.005   & 0.637 ± 0.011 \\
AB$'$C$'$D$'$ & 0.930  & 0.800 ± 0.0004  & 0.750 ± 0.016   & 0.804 ± 0.008   & 0.719 ± 0.022   & 0.770 ± 0.031 \\
A$'$BC$'$D    & 0.800  & 0.687 ± 0.0006  & 0.639 ± 0.015   & 0.683 ± 0.019   & 0.615 ± 0.028   & 0.665 ± 0.033 \\
A$'$B'CD$'$   & -0.525 & -0.450 ± 0.0003 & -0.427 ± 0.008  & -0.436 ± 0.038  & -0.411 ± 0.007  & -0.450 ± 0.008 \\
\hline
\textbf{Inequality} & 3.055  & 2.626 ± 0.001   & 2.472 ± 0.042   & 2.599 ± 0.073   & 2.370 ± 0.062   & 2.560 ± 0.080 \\
\hline
\end{tabular}

\caption{State Vector based method with all results in IBM kyiv using 21 operators}
\label{tab:state_dicke_kyiv}
\end{table*}

\begin{table*}[]

\begin{tabular}{|l|c|c|c|c|c|c|}
\hline
\textbf{Operator} & \textbf{Ideal} & \multicolumn{5}{c|}{\textbf{Statevector IBM\_Sherbrook}} \\
\cline{3-7}
                  &                & \textbf{Simulator (Noisy)} & \textbf{Without Mitigation} & \textbf{M3} & \textbf{DD+T} & \textbf{DD+T+M3} \\

\hline
ABCD              & 0.800   & 0.674 ± 0.007   & 0.682 ± 0.007   & 0.738 ± 0.004   & 0.597 ± 0.003   & 0.644 ± 0.006 \\
AB$'$C$'$D$'$     & 0.930   & 0.782 ± 0.012   & 0.794 ± 0.009   & 0.841 ± 0.012   & 0.696 ± 0.007   & 0.753 ± 0.007 \\
A$'$BC$'$D        & 0.800   & 0.674 ± 0.007   & 0.683 ± 0.007   & 0.728 ± 0.011   & 0.600 ± 0.006   & 0.648 ± 0.011 \\
A$'$B$'$CD$'$     & -0.525  & -0.445 ± 0.003  & -0.446 ± 0.004  & -0.487 ± 0.003  & -0.391 ± 0.011  & -0.419 ± 0.011 \\
\hline
\textbf{Inequality} & 3.055 & 2.576 ± 0.024   & 2.607 ± 0.029   & 2.599 ± 0.073   & 2.285 ± 0.019   & 2.466 ± 0.035 \\
\hline
\end{tabular}

\caption{State Vector based method with all results in IBM Sherbrooke using 21 operators}
\label{tab:state_dicke_sherbrook}
\end{table*}

\begin{figure}
    \centering
    \includegraphics[width=1\linewidth]{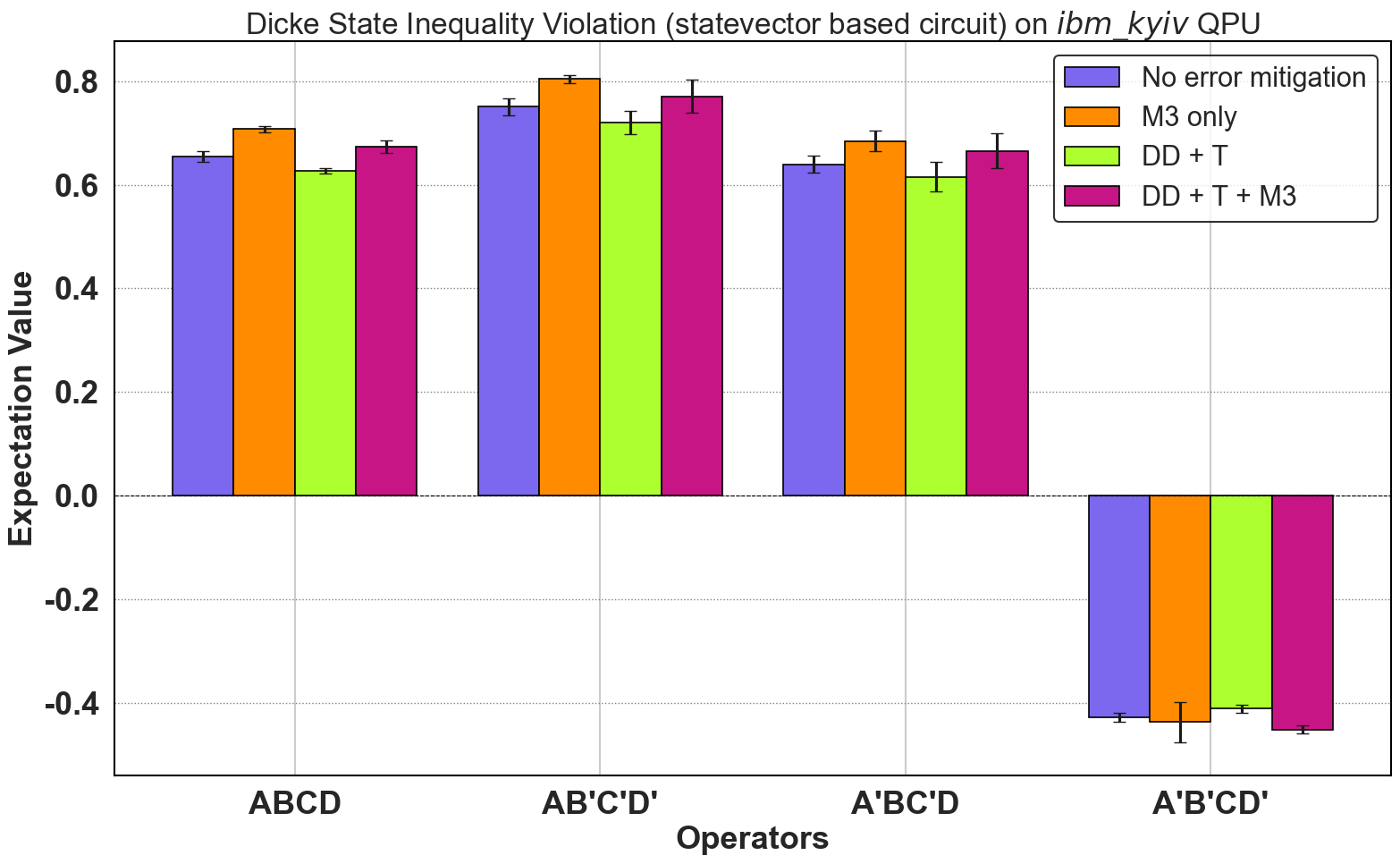}
    \caption{A comparison of the average expectation values of the measurement operator for the violation of the Bell-type inequality without and with various mitigation techniques on ibm\_kyiv, using the state vector method. }
    \label{state_four_op_Dicke_kyiv}
\end{figure}

\begin{figure}
    \centering
    \includegraphics[width=1\linewidth]{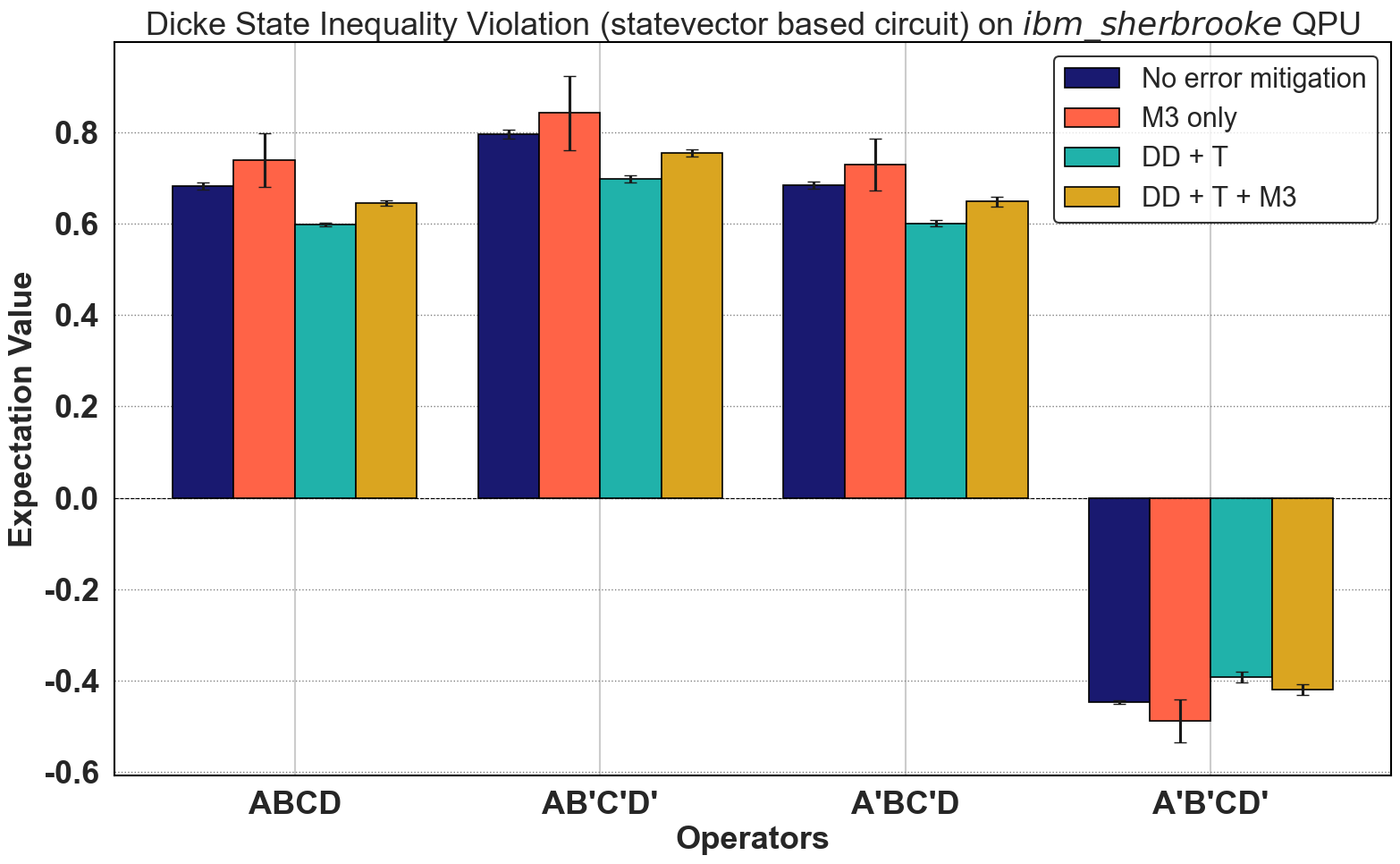}
    \caption{A comparison of the average expectation values of the measurement operator for the violation of the Bell-type inequality without and with various mitigation techniques on ibm\_sherbrook, using the state vector method. }
    \label{state_four_op_Dicke_sherbrook}
\end{figure}

For the Dicke state prepared by the state vector-based method, the experimental average values of all four operators and the corresponding Bell parameter are given in Table \ref{tab:state_dicke_kyiv} and Table \ref{tab:state_dicke_sherbrook}. Using this method, we observed from Tables~\ref{tab:state_dicke_kyiv} and \ref{tab:state_dicke_sherbrook} that a violation of the Bell-type inequality was achieved even without applying any error mitigation. This is likely because the state-vector-based approach prepares the Dicke state with inherently lower noise compared to the gate-based method. Interestingly, applying additional error mitigation techniques did not consistently lead to a linear improvement in the violation. In fact, the maximum violation was obtained using M3 on \texttt{ibm\_kyiv}, and without any mitigation on \texttt{ibm\_sherbrook}. Our analysis suggests that while the state prepared by this method is already less noisy, applying further advanced mitigation techniques can sometimes introduce additional distortions rather than reducing them. Similarly, comparison of the expectation values of all four measurements with different mitigation techniques using two IBM QPUs is given in the Figs.~\ref{state_four_op_Dicke_kyiv} and  \ref{state_four_op_Dicke_sherbrook}.

\begin{figure}
    \centering
    \includegraphics[width=1\linewidth]{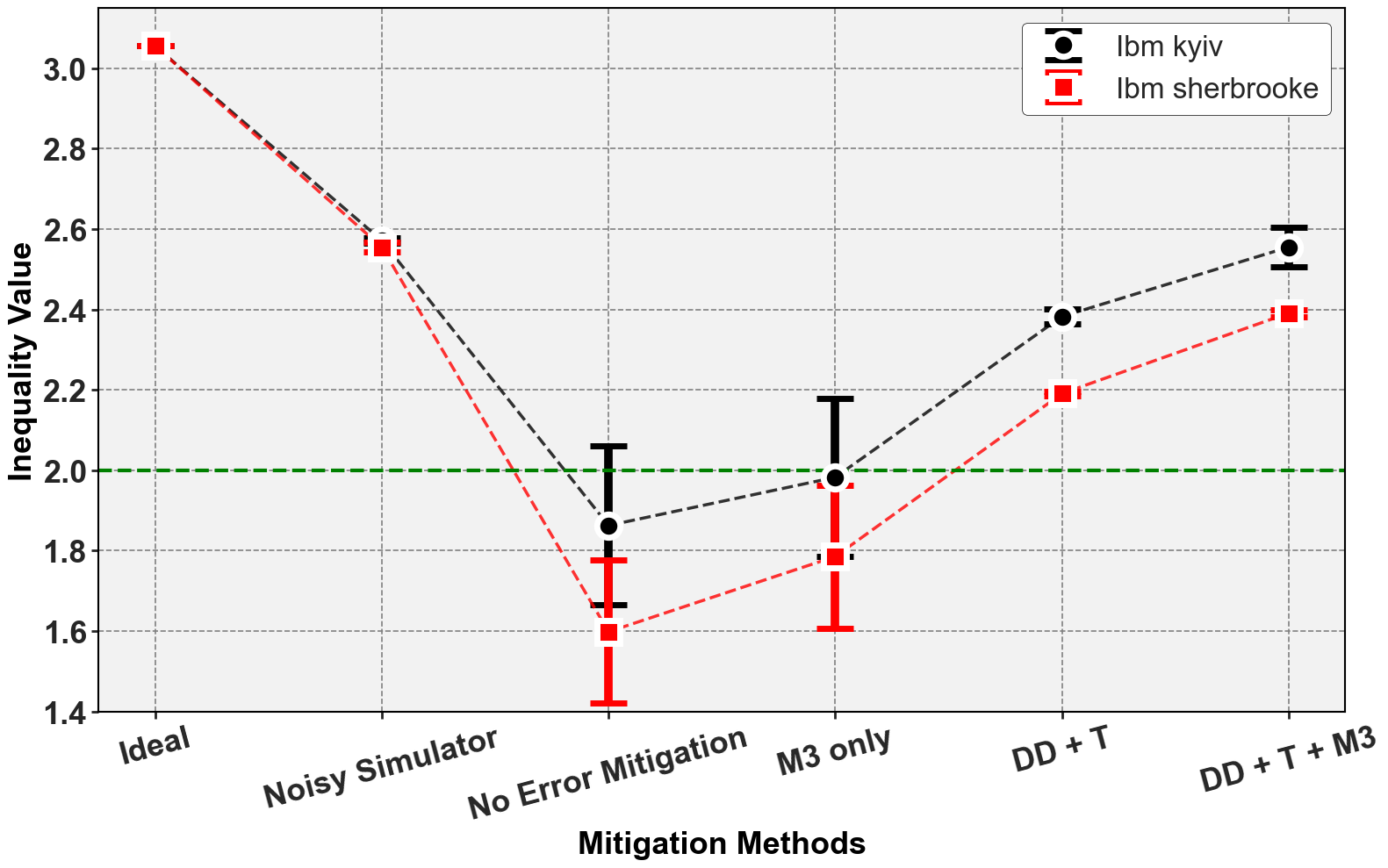}
    \caption{A comparison of the Bell type inequality violation values of Ideal Simulator, noisy, and real hardware with and without mitigation methods on two IBM QPUs for Dicke state prepared by gate-based method.}
    \label{gate_dicke_comparison}
\end{figure}

\begin{figure}
    \centering
    \includegraphics[width=1\linewidth]{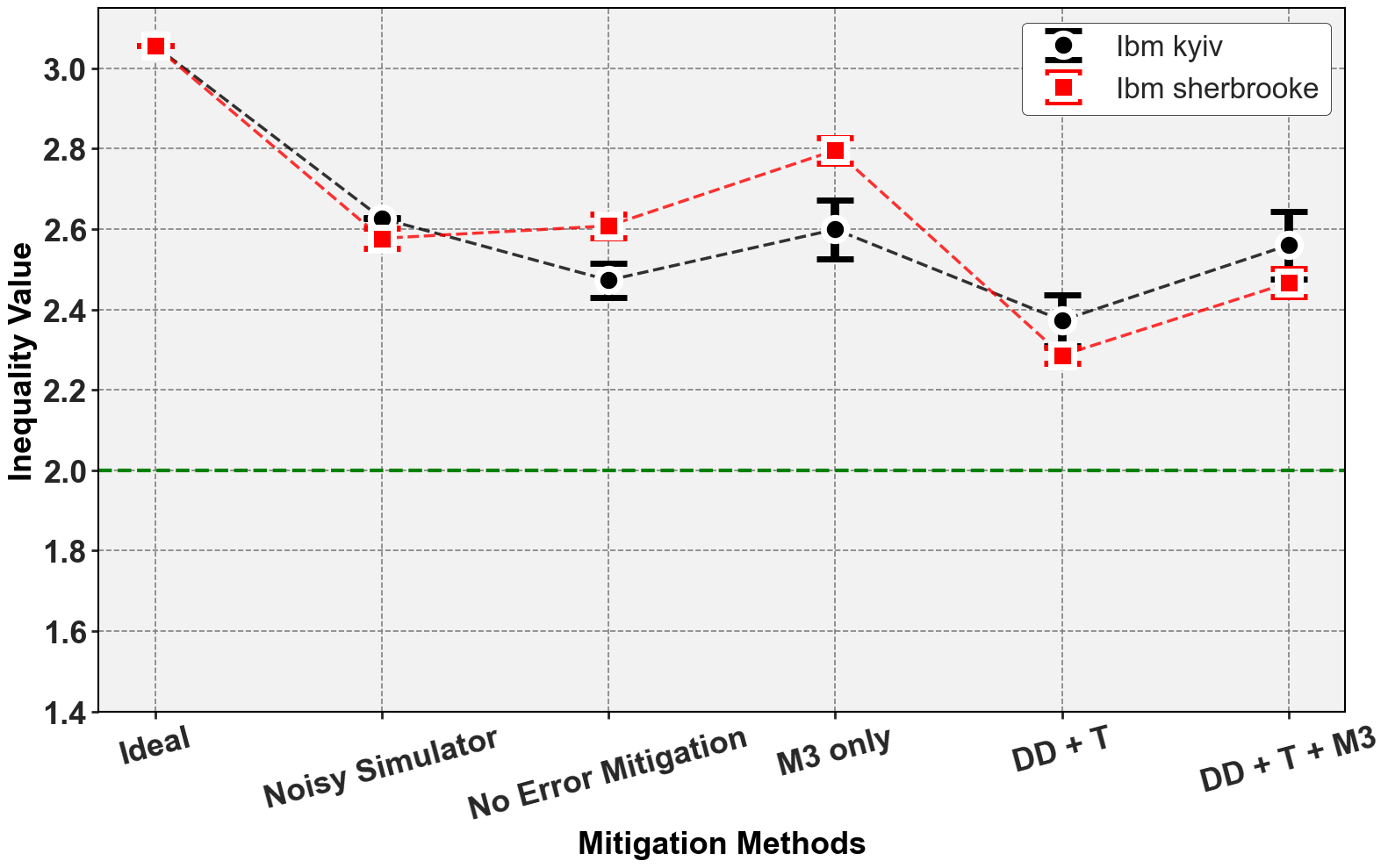}
    \caption{A comparison of the Bell type inequality violation values of Ideal Simulator, noisy, and real hardware with and without mitigation methods on two IBM QPUs for Dicke state prepared by statevector-based method.}
    \label{state_dicke_comparison}
\end{figure}

A comparison of Bell-type inequality values prepared using two methods and evaluated on the two different IBM QPUs, using various mitigation techniques, is provided in Fig. \ref{gate_dicke_comparison}, and \ref{state_dicke_comparison}. The state-vector-based method gives the best result of the Bell parameter as compared to the gate-based method. This could be mainly due to a difference in circuit depth; however, the state prepared by the gate-based method is not exact, while the one prepared by the statevector is exact.


 The difference in the result of noisy simulator and hardware without error mitigation can be due to the fact that, noisy simulator use snapshots of noise of real systems to model qubit properties like error rates and coherence times, they do not capture the dynamic nature of real-time fluctuations in hardware performance, such as drift in qubit calibration or environmental noise. Furthermore, while the noisy simulation can approximate certain errors, it cannot fully replicate the cumulative impact of hardware imperfections, thermal noise, and operational variability that arise during actual executions. Consequently, real hardware results typically experience higher error rates than those predicted by noisy simulations.

For the Bell state, the theoretical maximum violation is 2.82, with the experimental result closely matching at 2.81 using the M3 mitigation on ibm\_sherbrook. In contrast, for the Dicke state, the theoretical maximum is 3.055, but the experimental value reaches only 2.39 and 2.555 on ibm\_sherbrook and ibm\_kyiv for the gate-based method using DD+T+M3 mitigation and 2.599 on ibm\_kyiv using M3 mitigation and 2.607 on ibm\_sherbrook without error mitigation for the state vector-based method, respectively. This discrepancy can largely be attributed to the deeper circuit depths required for the Dicke state, with depths of 143 and 126, respectively, for Dicke state preparation by gate-based method and state vector-based method,  compared to just 9 for the Bell state. The greater depth increases the accumulation of gate and decoherence errors. Additionally, the Bell state requires only 3 measurements to test its inequality, while the Dicke state involves 21 measurements. Each measurement introduces potential errors, including noise and decoherence, which accumulate across all measurements, contributing significantly to the observed deviation from the theoretical maximum in the Dicke state case.

We also investigated the Bell-type inequality violation using all 81 possible operator combinations on IBM QPUs for the Dicke state prepared via both the gate-based and statevector-based methods. For the gate-based preparation, we observed no violation, with Bell parameter values of 1.31 and 1.25 on the respective QPUs. These measurements were performed without error mitigation due to the 10-minute execution time limitation on IBM's free-tier access. Applying error mitigation techniques in this case would exceed the time constraints, as multiple runs are required to obtain statistically significant results. For the Dicke state prepared using the statevector method, we similarly did not observe any violation. 

\section{conclusion}
In conclusion, we have designed a customized operator that can violate any Bell-type inequalities for any entangled state. Our customized operator is the sum of Pauli matrices ($\sigma_x$, $\sigma_y$, and $\sigma_z$). We investigated the violation of Bell-type inequalities theoretically and experimentally for two- and four-qubit systems on various IBM QPUs. For the two-qubit Bell state, we got the violation without and with error mitigation method on both the IBM QPUs; however, M3 error mitigation on ibm\_sherbrook significantly enhances the observed Bell violations, bringing them closer to theoretical predictions.  By implementing both gate-based and statevector-based preparation methods for the Dicke state, we analyzed the effectiveness of different error mitigation strategies. Our results demonstrate that advanced mitigation techniques improved the Bell violation for the Dicke state prepared by the gate-based method. For the statevector-based method, violation was observed even without mitigation, suggesting the generation of a relatively low-noise state. 
However, our findings also reveal that increasing the complexity of mitigation does not always lead to linear improvements, especially when the prepared state is already near optimal. Additionally, we found no violation using all 81 operator combinations, highlighting the limitations imposed by current hardware noise and time constraints on public quantum systems. In our investigation, the state vector-based method gives better results than the gate-based method. This could be because it initializes the system directly into the desired Dicke state, minimizing the overall circuit depth and noise exposure. By bypassing intermediate gate operations, this approach reduces the likelihood of errors from decoherence and gate imperfections. Even though the gate-based method may use fewer CNOT gates, the statevector method's combination of lower circuit depth, fewer intermediate operations, and potentially more optimized qubit routing leads to better results on real hardware.

This study underscores the importance of tailoring mitigation techniques to specific state preparation methods and opens avenues for further exploration of noise-resilient quantum protocols on near-term devices. We have successfully shown the violation of the Bell-type inequalities for the two-qubit Bell state and four-qubit Dicke state using our customized operator. In the future, that customized operator can be applied to any Bell-type inequality to any number of entangled states.

\section*{Data Availability statement}
All data that support the findings of this study are included in the article.
 
\section*{Acknowledgement}
VN acknowledges the support from the Interdisciplinary Cyber Physical Systems (ICPS) programme of the Department of Science and Technology (DST), India, Grant No.: DST/ICPS/QuST/Theme-1/2019/6.
Tomis Prajapati would like to acknowledge CSIR for the research funding.

\begin{widetext}
   
\appendix

\section*{Appendix}

\section{Derivation of the Bell polynomial for two-qubit Bell state}
Here, we derive the Bell polynomial for the two-qubit Bell state \(\vert \Phi^+ \rangle\), as described in Eq.~(\ref{Eq3}). Initially, we compute the first term of Eq.~(\ref{Eq3}), as shown in Eq.~(\ref{Eq4}). Following a similar approach, we calculate all the remaining terms. Consequently, the Bell polynomial can be expressed as follows:

\begin{equation}
\label{Eq5}
\begin{aligned}
\left| S_{|\phi^+\rangle} \right| &= 
\cos(\theta_{A'}) \left(-\cos(\theta_B) + \cos(\theta_{B'})\right) + \cos(\theta_A) \left(\cos(\theta_B) + \cos(\theta_{B'})\right) + \cos(\phi_A + \phi_B) \sin(\theta_A) \sin(\theta_B)
\\&- \cos(\phi_{A'} + \phi_B) \sin(\theta_{A'}) \sin(\theta_B) + \cos(\phi_A + \phi_{B'}) \sin(\theta_A) \sin(\theta_{B'})
+ \cos(\phi_{A'} + \phi_{B'}) \sin(\theta_{A'}) \sin(\theta_{B'}).
\end{aligned}
\end{equation}

\section{Derivation of the Bell polynomial for four-qubit Dicke state}
Here, we took a similar approach to that taken for the two-qubit Bell state.
Eq. (\ref{eq1}) expressed the four-qubit Dicke state ($|D^{(2)}_4\rangle$), and its Bell-type inequality is defined in Eq. (\ref{dicke_bell}). For calculating the Bell polynomial, we first calculate the 1st term  in Eq. (\ref{dicke_bell}) as follows;

\begin{multline}
\label{b2}
\langle\hat{A}\hat{B}\hat{C}\hat{D}\rangle =
\frac{1}{3} \Bigg(
\sin(\theta_A) \Big[
-2 \cos(\theta_C) \Big( \cos(\theta_D) \cos(\phi_A - \phi_B) \sin(\theta_B)
+ \cos(\theta_B) \cos(\phi_A - \phi_D) \sin(\theta_D) \Big) \\
+ \sin(\theta_C) \Big(
-2 \cos(\theta_B) \cos(\theta_D) \cos(\phi_A - \phi_C)
+ \big( \cos(\phi_A + \phi_B - \phi_C - \phi_D)
+ 2 \cos(\phi_A - \phi_B) \cos(\phi_C - \phi_D) \big)
\sin(\theta_B) \sin(\theta_D) \Big)
\Big] \\
+ \cos(\theta_A) \Big[
-2 \sin(\theta_B) \Big( \cos(\theta_D) \cos(\phi_B - \phi_C) \sin(\theta_C)
+ \cos(\theta_C) \cos(\phi_B - \phi_D) \sin(\theta_D) \Big) \\
+ \cos(\theta_B) \Big(
3 \cos(\theta_C) \cos(\theta_D)
- 2 \cos(\phi_C - \phi_D) \sin(\theta_C) \sin(\theta_D)
\Big)
\Big]
\Bigg)
.
\end{multline}

\section{Expectation value calculation}
After executing the quantum circuits on the real IBM QPU, the results are returned as a set of measurement counts. These counts represent how many times each possible outcome (bitstring) is measured in the given number of shots (repetitions of the circuit). For a two-qubit system, the possible outcomes are the bitstrings `00`, `01`, `10`, and `11`, corresponding to the measurement of the computational basis states $\vert 00\rangle$, $\vert 01\rangle$, $\vert 10\rangle$, and $\vert 11\rangle$, respectively.

\subsubsection{Mapping Counts to Expectation Values}

To calculate the expectation value of an operator $X$, we use the measured counts to compute the probability of each outcome. The expectation value is then given by:

\[
\langle X \rangle = \sum_{\text{outcomes}} P(\text{outcome}) \cdot \text{eigenvalue(outcome)}
\]

where $P(\text{outcome})$ is the probability of measuring a specific bitstring, and the corresponding eigenvalue depends on the operator and the outcome. For Pauli operators, the eigenvalues are $\pm 1$, as each Pauli matrix has eigenvalues in this range when measured in its eigenbasis.

For example, when measuring the Pauli-$Z$ operator on a qubit, the outcomes `0` and `1` correspond to eigenvalues $+1$ and $-1$, respectively. Thus, the expectation value of $Z$ is calculated as:
\[
\langle Z \rangle = P(0) \cdot (+1) + P(1) \cdot (-1) = P(0) - P(1)
\]
where $P(0)$ and $P(1)$ are the probabilities of measuring `0' and `1', respectively.

In the case of two-qubit operators like $\sigma_x \otimes \sigma_x$, the outcomes correspond to eigenvalues derived from the product of the eigenvalues of the individual Pauli operators acting on each qubit. For instance, the outcome `00' would correspond to eigenvalue $+1 \cdot +1 = +1$, while the outcome `11' would correspond to $-1 \cdot -1 = +1$. For each operator, we map each possible bitstring to its corresponding eigenvalue, then compute the expectation value as a weighted sum over the measurement probabilities.






\subsubsection{Summing Contributions from Decomposed Operators}

The expectation value for each operator is then calculated as described above, using the mitigated counts to compute the probabilities of each outcome. For example, if we measure the operator $\sigma_x \otimes \sigma_x$ and obtain corrected counts \{‘00’: 5000, ‘01’: 0, ‘10’: 0, ‘11’: 5000\} out of 10,000 shots, the expectation value is:

\[
\langle XX \rangle = \frac{5000}{10000} \cdot (+1) + \frac{5000}{10000} \cdot (+1) = 1
\]

This process is repeated for all operators of interest, such as $\sigma_y \otimes \sigma_y$ and $\sigma_z \otimes \sigma_z$, allowing us to compute their respective expectation values.

As we have decomposed the operators $A$, $B$, $A'$, and $B'$ into linear combinations of Pauli operators, the overall expectation value of an operator like $  A\otimes B$ is a weighted sum of the expectation values of the individual Pauli operators. For instance, if:
\[
A =  a_1 \sigma_x + a_2 \sigma_y + a_3 \sigma_z
\]
and similarly for $B$, the expectation value of $A \otimes B$ is:
\[
\langle A \otimes B \rangle = a_1 b_1 \langle XX \rangle + a_2 b_2 \langle YY \rangle + a_3 b_3 \langle ZZ \rangle
\]
where $\langle XX \rangle$, $\langle YY \rangle$, and $\langle ZZ \rangle$ are the expectation values of the Pauli operators measured on the Bell state. This method allows us to compute the overall expectation value for any operator expressed in terms of Pauli matrices, leveraging that only a small subset of operators have non-zero contributions.

\section{Calibration Data of utilized QPUs}

Calibration data of \textit{ibm\_kyiv} obtained from the IBM quantum platform during one of the executions on 13 April 2025 at 02:25:26 UTC. In the case of the four-qubit Dicke state, the utilized qubits were q\_110, q\_117, q\_118, and q\_119, while for the Bell State, the utilized qubits were q\_118 and q\_119.
\begin{table*}[ht]
\centering
\begin{tabular}{|c|c|c|c|c|c|l|}
\hline
\textbf{Qubit} & \textbf{T1 ($\mu$s)} & \textbf{T2 ($\mu$s)} & \textbf{Frequency (GHz)} & \textbf{Readout error} & \textbf{Pauli-X error} & \textbf{ECR error} \\
\hline
q\_110 & 192.0918288 & 77.81293221 & 4.805827445 & 0.01147460938 & 0.000213357699 & ECR(110\_118): 0.013833191509604048 \\
\hline
q\_117 & 233.7851969 & 57.49821207 & 4.831597205 & 0.02587890625 & 0.0004991485813 & ECR(117\_118): 0.009432063277649116 \\
\hline
q\_118 & 183.4727162 & 211.8473848 & 4.749114666 & 0.003662109375 & 0.0004475850526 & 
\makecell[l]{ECR(118\_117): 0.009432063277649116 \\ 
ECR(118\_119): 0.0123333631852508 \\ 
ECR(118\_110): 0.013833191509604048} \\
\hline
q\_119 & 361.1192799 & 317.4148785 & 4.610730335 & 0.00244140625 & 0.0001937314091 & ECR(119\_118): 0.0123333631852508 \\
\hline
\end{tabular}
\caption{Calibration data from \textit{ibm\_kyiv} for Physical Qubits used for execution of Inequality for Bell State and Dicke State. Including relaxation and dephasing times, frequencies, readout errors, Pauli-X gate errors, and ECR gate errors.}
\label{tab:calibration_kyiv}
\end{table*}

Calibration data of \textit{ibm\_sherbrooke} obtained from the IBM quantum platform during one of the executions on 08 April 2025 at 07:16:32 UTC. In the case of the four-qubit Dicke state, the utilized qubits were q\_69, q\_70, q\_74, and q\_89, while for the Bell State, the utilized qubits were q\_70 and q\_74.

\begin{table}[h]
\centering
\begin{tabular}{|c|c|c|c|c|c|l|}
\hline
\textbf{Qubit} & \textbf{T1 ($\mu$s)} & \textbf{T2 ($\mu$s)} & \textbf{Frequency (GHz)} & \textbf{Readout error} & \textbf{Pauli-X error} & \textbf{ECR error} \\
\hline
q\_69 & 306.3696555 & 30.43884575 & 4.837439034 & 0.00830078125 & 0.0001885654664 & ECR(69\_70): 0.00562455142472329 \\
\hline
q\_70 & 367.1761525 & 426.2109472 & 4.70854747 & 0.0087890625 & 0.0001389173346 & 
\makecell[l]{ECR(70\_69): 0.00562455142472329 \\ 
ECR(70\_74): 0.0045853921760504734} \\
\hline
q\_74 & 172.1878141 & 253.7196114 & 4.8143782 & 0.0029296875 & 0.000213256068 & 
\makecell[l]{ECR(74\_70): 0.0045853921760504734 \\ 
ECR(74\_89): 0.007157467756690794} \\
\hline
q\_89 & 290.1616709 & 377.5089279 & 4.947939584 & 0.01806640625 & 0.0001196076968 & ECR(89\_74): 0.007157467756690794 \\
\hline
\end{tabular}
\caption{Calibration data from \textit{ibm\_sherbrooke} for Physical Qubits used for execution of Inequality for Bell State and Dicke State. Including relaxation and dephasing times, frequencies, readout errors, Pauli-X gate errors, and ECR gate errors.}
\label{tab:calibration_sherbrooke}
\end{table}

\section{Comparison of obtained counts}
An inequality test of the four-qubit Dicke state was performed with no mitigation, with M3 mitigation only, DD + T, and DD + T + M3. The difference in obtained counts when the circuit is prepared using the unitary gate-based method for different operators is shown in Fig. \ref{fig:comparison_counts_Kyiv_gate}. 

\begin{figure}[htbp]
    \centering
    \subfigure[]{
        \includegraphics[width=0.3\textwidth]{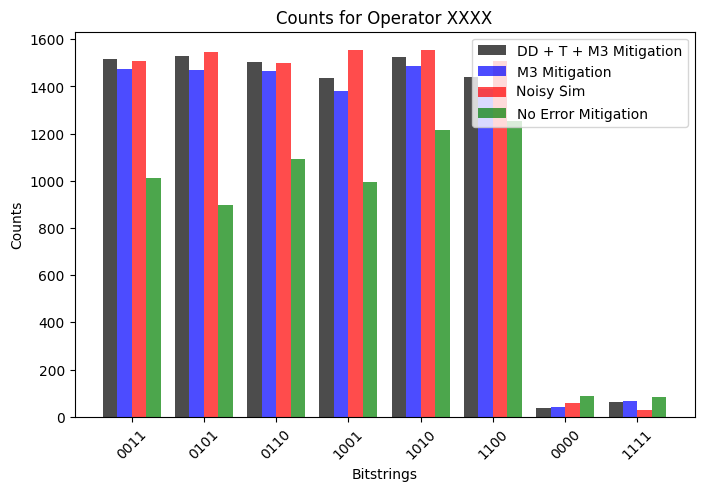}
        \label{fig:fig1}
    }
    \hfill
    \subfigure[]{
        \includegraphics[width=0.3\textwidth]{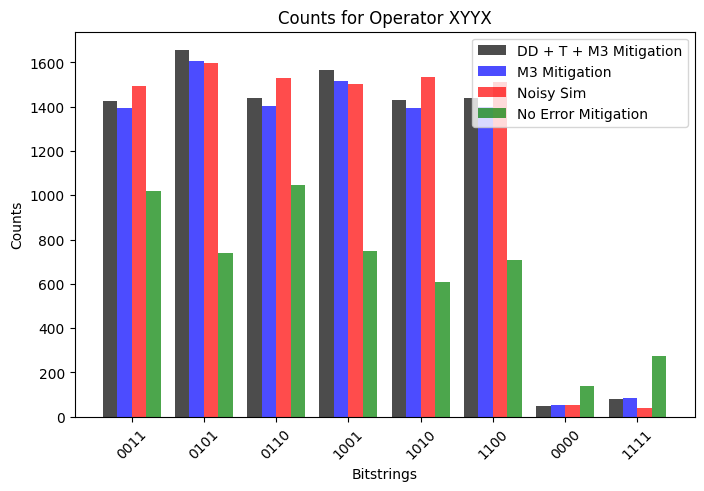}
        \label{fig:fig2}
    }
    \hfill
    \subfigure[]{
        \includegraphics[width=0.3\textwidth]{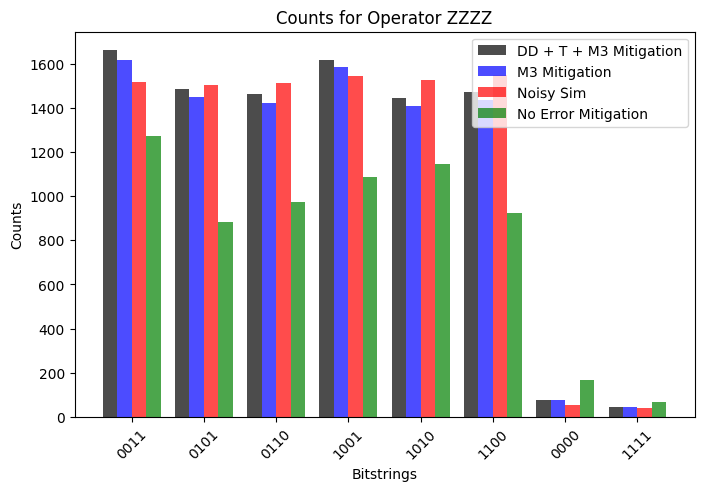}
        \label{fig:fig2}
    }
    \caption{A comparison of obtained counts for operator \textit{XXXX}, \textit{XYYX}, and \textit{ZZZZ} with Different mitigation and no mitigation on ibm\_kyiv and noisy simulator, with circuit preparation method being Unitary Gate Based Method for Operator.}
    \label{fig:comparison_counts_Kyiv_gate}
\end{figure}

Similarly, the difference in obtained counts when the circuit is prepared using the statevector-based method for different operators is shown in Fig. \ref {fig:comparison_counts_Kyiv_statevector}.

\begin{figure}[htbp]
    \centering
    \subfigure[]{
        \includegraphics[width=0.3\textwidth]{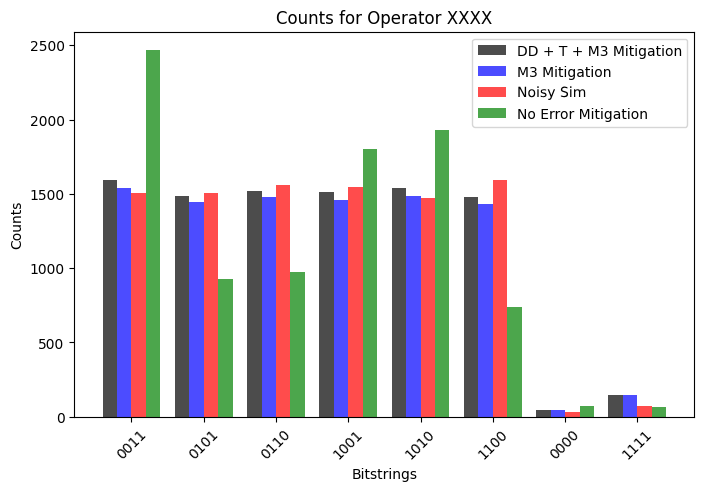}
        \label{fig:fig1}
    }
    \hfill
    \subfigure[]{
        \includegraphics[width=0.3\textwidth]{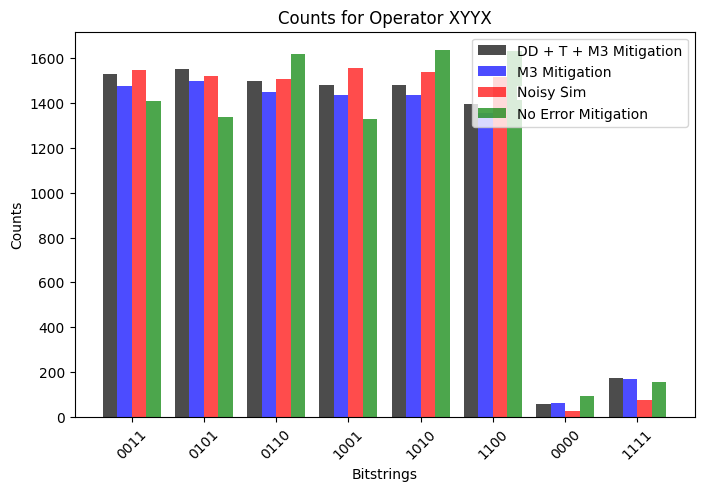}
        \label{fig:fig2}
    }
    \hfill
    \subfigure[]{
        \includegraphics[width=0.3\textwidth]{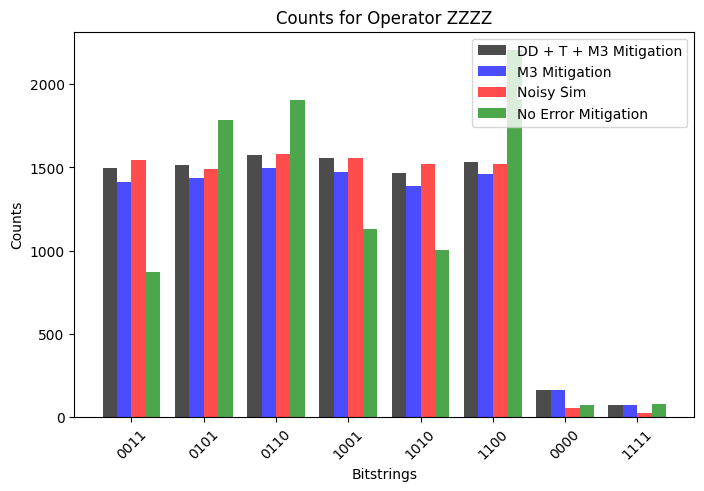}
        \label{fig:fig2}
    }
    \caption{A comparison of obtained counts for operator \textit{XXXX}, \textit{XYYX}, and \textit{ZZZZ} with Different mitigation and no mitigation on ibm\_kyiv and noisy simulator, with circuit preparation method being statevector based method for Operator.}
    \label{fig:comparison_counts_Kyiv_statevector}
\end{figure}
\end{widetext}

\bibliography{ref}

\end{document}